\documentclass[preprint]{aastex6}
\def\degpoint{\ifmmode ^{\rm{o}}\!. \else $^{\rm{o}}\!.$\fi}

\newcommand{\ms}{\mbox{m\,s$^{-1}$}}
\newcommand{\kms}{\mbox{km\,s$^{-1}$}}
\newcommand{\Msun}{\mbox{$M_{\odot}$}}
\newcommand{\Rsun}{\mbox{$R_{\odot}$}}
\newcommand{\Mjup}{\mbox{$M_{\rm Jup}$}}

\newcommand{\Lsun}{\mbox{$L_{\odot}$}}

\newcommand{\gtsimeq}{\raisebox{-0.6ex}{$\,\stackrel
         {\raisebox{-.2ex}{$\textstyle >$}}{\sim}\,$}}

\usepackage{graphicx}
\usepackage{subfigure}

\begin{document}

\title{The Pan-Pacific Planet Search VI: Giant planets orbiting 
HD\,86950 and HD\,222076 }

\author{Robert A.~Wittenmyer\altaffilmark{1,2}, M.I. 
Jones\altaffilmark{3}, Jinglin Zhao\altaffilmark{2}, J.P. 
Marshall\altaffilmark{2}, R.P. Butler\altaffilmark{4}, C.G. 
Tinney\altaffilmark{2}, Liang Wang\altaffilmark{5}, John Asher 
Johnson\altaffilmark{6} }

\altaffiltext{1}{Computational Engineering and Science Research Centre, 
University of Southern Queensland, Toowoomba, Queensland 4350, 
Australia}
\altaffiltext{2}{School of Physics and Australian Centre for 
Astrobiology, UNSW Australia, Sydney 2052, Australia}
\altaffiltext{3}{Department of Electrical Engineering and Center of 
Astro-Engineering UC, Pontificia Universidad Cato\'lica de Chile, Av. 
Vicuna Mackenna 4860, 782-0436 Macul, Santiago, Chile}
\altaffiltext{4}{Department of Terrestrial Magnetism, Carnegie 
Institution of Washington, 5241 Broad Branch Road, NW, Washington, DC 
20015-1305, USA}
\altaffiltext{5}{Key Laboratory of Optical Astronomy, National 
Astronomical Observatories, Chinese Academy of Sciences, A20 Datun Road, 
Chaoyang District, Beijing 100012, China}
\altaffiltext{6}{Harvard-Smithsonian Center for Astrophysics, Cambridge, 
MA 02138 USA}
\email{rob@unsw.edu.au}

\shorttitle{PPPS VI: Two new giant planets }
\shortauthors{Wittenmyer et al.}


\begin{abstract}

\noindent We report the detection of two new planets orbiting the K 
giants HD\,86950 and HD\,222076, based on precise radial velocities 
obtained with three instruments: AAT/UCLES, FEROS, and CHIRON.  
HD\,86950b has a period of 1270$\pm$57 days at $a=2.72\pm$0.08\,AU, and 
m sin $i=3.6\pm$0.7\,\Mjup.  HD\,222076b has $P=871\pm$19 days at 
$a=1.83\pm$0.03\,\,AU, and m sin $i=1.56\pm$0.11\,\Mjup.  These two 
giant planets are typical of the population of planets known to orbit 
evolved stars.  In addition, we find a high-amplitude periodic velocity 
signal ($K\sim$50\,\ms) in HD\,29399, and show that it is due to stellar 
variability rather than Keplerian reflex motion.  We also investigate 
the relation between planet occurrence and host-star metallicity for the 
164-star Pan-Pacific Planet Search sample of evolved stars.  In spite of 
the small sample of PPPS detections, we confirm the trend of increasing 
planet occurrence as a function of metallicity found by other studies of 
planets orbiting evolved stars.

\end{abstract}



\section{Introduction}

Nearly all our knowledge of planets orbiting stars more massive than the 
Sun comes from Doppler surveys targeting these stars after they have 
evolved off the main sequence.  Subgiants and first-ascent giants are 
suitable radial velocity targets because their surface gravities remain 
high enough (log $g\gtsimeq$3) to avoid the large-amplitude pulsations 
common in red giants \citep{hekker08}.  Many groups have been conducting 
precise radial velocity surveys of these stars, with $\sim$1000 total 
targets and 10-15 years of observations \citep[e.g.][]{sato05, 
johnson06b, reffert15}.

The Pan-Pacific Planet Search (PPPS - Wittenmyer et al.~2011b) operated 
on the 3.9m Anglo-Australian Telescope (AAT) between 2009 and 2015, 
surveying 164 southern giant stars in search of planets as a southern 
extension of the Northern ``retired A stars'' programme 
\citep{johnson06b}.  Since the conclusion of AAT observations in 2014, 
we have published a series of planet discoveries combining PPPS 
observations with data from the Okayama planet search \citep{sato13, 
47366} and the Chile-based EXPRESS survey, with which we share 37 
targets \citep{121056, 155233, 33844, jones16}.

We have secured enough detections from the PPPS sample to make 
quantitative statements about the occurrence rate of giant planets 
orbiting evolved stars.  For dwarf stars, the metallicity [Fe/H] is now 
well-known to be positively correlated with the probablity of hosting a 
giant planet \citep[e.g.][]{gonz97, santos01, fv05}.  This is a 
consequence of the core-accretion model of planet formation, wherein 
metal-rich disks are more efficient at forming cores due to their 
enhanced surface density of solids \citep{pollack96}.

For giant stars, however, there remains disagreement as to the 
presence of such a planet-metallicity correlation.  \citet{hekker07} 
found that planet-hosting giants had metallicities 0.13$\pm$0.03 dex 
higher than the overall sample of 380 G and K giants.  Their conclusion 
was bolstered by a subsequent analysis of the same sample with a further 
seven years of observational data \citep{reffert15}.  But 
\citet{takeda08} found no difference in the metallicities of planet 
hosts versus non-hosts in their 322-star Okayama Planet Search sample.  
A preliminary analysis of the Penn State-Torun Planet Search sample by 
\citet{z10} hinted at an anticorrelation, with the metallicities of 
RV-variable giant stars 0.15$\pm$0.06 dex lower than the RV-stable 
stars.  \citet{maldonado13} and \citet{mortier13} analysed large samples 
of giant stars, derived stellar parameters in a homogeneous manner, and 
found no significant metallicity differences between stars hosting 
planets and those not.  More recently, however, \citet{jones16} 
presented an analysis of 166 giant stars and found a distinct peak in 
planet occurrence at [Fe/H]$\sim\,+0.35$.  Clearly, the issue is far 
from settled. 

In this paper, we present two new giant planets from the PPPS sample, in 
conjunction with data from the EXPRESS survey \citep{jones11}.  Section 
2 briefly describes the observational data and host star properties.  
Section 3 gives the results of the orbit fitting and describes the 
vetting process, and Section 4 places these planetary systems in context 
and discusses the planet-metallicity correlation within the PPPS sample.


\section{Observations and Stellar Properties}
\label{obs}

Observations have been obtained with three instruments in two parallel 
planet-search efforts using the UCLES spectrograph \citep{diego90} on 
the AAT, the CHIRON spectrograph \citep{toko13} on the 1.5m telescope at 
CTIO, and the FEROS spectrograph on the on the 2.2m telescope at La 
Silla \citep{kaufer99}.

UCLES delivers a resolution of $\lambda$/$\Delta\lambda$$\approx$45,000 
with a 1 arcsecond slit, on which an iodine cell imprints a dense forest 
of narrow absorption lines used to calibrate the instrumental 
point-spread function is calibrated \citep{val:95,BuMaWi96}.  
FEROS on the 2.2m telescope at La Silla has a resolving power of 
$\sim$ 48,000, and uses a ThArNe lamp for precise wavelength 
calibration.  The instrument is equipped with a simultaneous calibration 
fibre which is used to correct the night spectral drift.  The FEROS 
radial velocities were computed using the simultaneous calibration 
method \citep{baranne96}, using an improved reduction code presented in 
Jones et al. (2016; submitted).  CHIRON is fed by a single fibre and an 
image slicer, delivering high resolution (R $\sim$ 80,000) and higher 
efficiency than the slit and narrow slit modes, also available.  The 
CHIRON echelle spectra are comprised of 71 orders covering a wavelength 
range between $\sim$ 4100-8800 \AA.  The spectrograph is also equipped 
with an iodine cell, which superimposes a rich absorption spectrum in 22 
different orders covering between $\sim$ 5000-6200 \AA. The radial 
velocities are computed with the iodine cell technique \citep{BuMaWi96}, 
following the newest reduction code described in Jones et al. (2016; 
submitted).  A summary of the observations is given in 
Table~\ref{obslog}, and the velocities from UCLES, FEROS, and CHIRON are 
given in Tables~2-4. 

Stellar properties were derived from iodine-free template UCLES spectra 
with $R\sim\,60,000$ as described fully in \citet{ppps5}.  In brief, 
spectroscopic stellar parameters were determined via a standard 1D, 
local thermodynamic equilibrium (LTE) abundance analysis using the 2013 
version of MOOG \citep{sne73} with the ODFNEW grid of Kurucz ATLAS9 
model atmospheres \citep{cas03}.  Complete stellar parameters from 
\citet{ppps5} and other literature sources are given in 
Table~\ref{stellarparams}.


\section{Orbit Fitting and Companion Parameters}

Although the data for these stars are sparse and poorly sampled, 
inspection suggested the presence of velocity signals, belied by the 
higher than usual scatter.  Following a well-trodden path 
\citep[e.g.][]{tinney11, 121056, 155233}, we first used a genetic 
algorithm to search a wide range of orbital periods, running for 10,000 
iterations (about $10^6$ possible configurations).  The period ranges 
were chosen based on visual inspection of the velocity data and are 
as follows: HD\,29399 -- [700-1100d]; HD\,86950 -- 
[1100-1400d]; HD\,222076 -- [800-1100d].  In all cases, convergence 
occurred rapidly, a hallmark of a genuine signal.  Again as in our 
previous work, we then used the best solution from the genetic algorithm 
as a starting point for fitting with a Keplerian model in the 
\textit{Systemic Console} version 2.1730 \citep{mes09}.  For all orbit 
fitting, 7\,\ms\ of jitter has been added in quadrature to the internal 
instrumental uncertainties of each data set.  This estimate is derived 
from the velocity scatter of 37 stable stars in the PPPS as first 
described in \citet{33844}.  Finally, we estimated the parameter 
uncertainties using the bootstrap routine within \textit{Systemic} on 
10,000 synthetic data set realisations.  The results are given in 
Table~\ref{planetparams}.

\subsection{HD 29399: A false positive}

\begin{figure}
\includegraphics[scale=0.4]{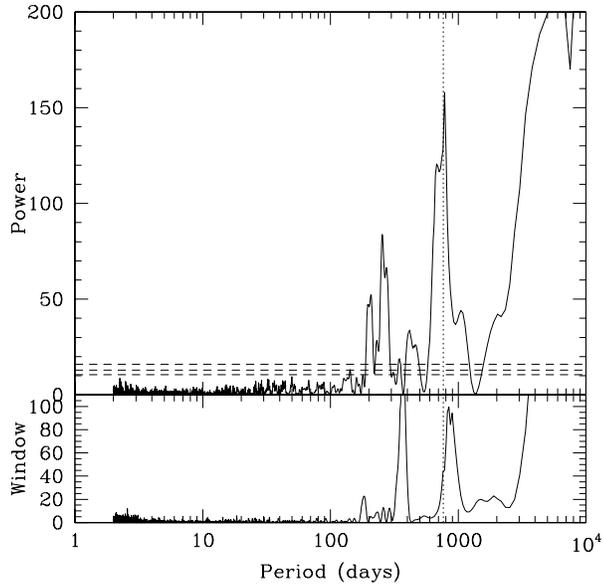}
\caption{Generalised Lomb-Scargle periodogram of ASAS photometry for 
HD\,29399.  A highly significant peak is present at the period of the 
radial velocity signal (765 days, dotted line).  10\%, 1\%, and 0.1\% 
bootstrap-derived false alarm probabilities are shown as horizontal 
dashed lines. }
\label{hd29399_phot}
\end{figure}

We find that the velocities for HD\,29399 can be fit with a planet 
having $P=765\pm$15 days and $K=52\pm$34\,\ms; the large uncertainty can 
be attributed to phase gaps which allow for a family of 
high-eccentricity solutions.  As in our previous work, we checked the 
All-Sky Automated Survey (ASAS) $V$ band photometric data \citep{asas} 
for variability due to intrinsic stellar processes.  We analysed a total 
of 984 epochs spanning 8.36 years, with a mean value of 6.07$\pm$0.32 
mag.  This scatter is an order of magnitude larger than that found for 
previously confirmed planet hosts from the PPPS \citep{121056, 33844}.  
The generalised Lomb-Scargle periodogram of this photometry \citep{zk09} 
is shown in Figure~\ref{hd29399_phot}, with the period of the candidate 
planet marked as a vertical dashed line.  There is an extremely 
significant periodicity squarely at the $\sim$765 day period of our 
radial velocities.  Given this evidence and the large photometric 
variability, we must conclude that the $K\sim$50\,\ms\ signal in the 
radial velocities is intrinsic to the star and not due to an orbiting 
planet.  While stellar rotation is frequently the culprit in 
radial velocity false positives \citep[e.g.][]{robertson15, rajpaul16, 
johnson16}, it is extremely unlikely in the case of HD\,29399.  The 
projected rotational velocity determined by \citet{med14} is less than 
1.2\,\kms; using the stellar radius of 4\,\Rsun\ in 
Table~\ref{stellarparams}, this gives a rotation period of 169 days.  Of 
course, since the measured v~sin~$i$ is a lower bound, it remains 
possible that the true rotational velocity is smaller, arising from a 
nearly pole-on orientation.  However, such an orientation would require 
an unreasonably high spot coverage to produce the observed variations in 
the radial velocities.  In Section 4.1, we note that HD\,29399 may host 
a debris disk; one might imagine quasi-periodic transits by debris 
\citep[e.g.][]{vanderburg15, croll15, rappaport16} to cause the 
photometric and radial velocity variations much as starspots would.  
However, we show in Section 4.1 that the required debris is best 
modelled at 1500\,K, and hence must orbit far too close to the star to 
produce a 765-day period by Keplerian orbital motion.  Indeed, for such 
a long periodicity, the most likely source is a stellar magnetic 
activity cycle.  In Section 3.3, we give further evidence via an 
analysis of the H$\alpha$ feature.

\subsection{Single planets orbiting HD 86950 and HD 222076}

\begin{figure}
\includegraphics[scale=0.6,angle=270]{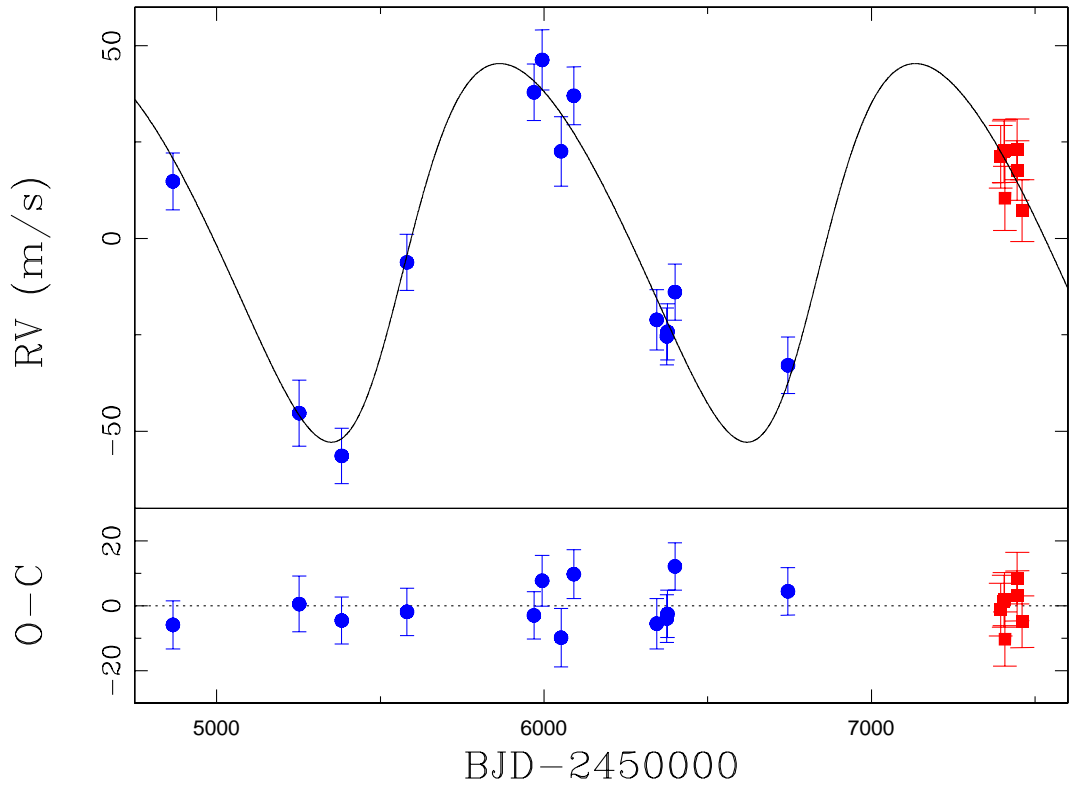}
\caption{Data and Keplerian fit for a 3.6\Mjup\ planet orbiting 
HD\,86950 (AAT -- blue, CHIRON -- red).  Error bars include 7\,\ms\ of 
jitter added in quadrature. The rms about this fit is 6.1\,\ms. }
\label{86950fit}
\end{figure}

\begin{figure}
\includegraphics[scale=0.4]{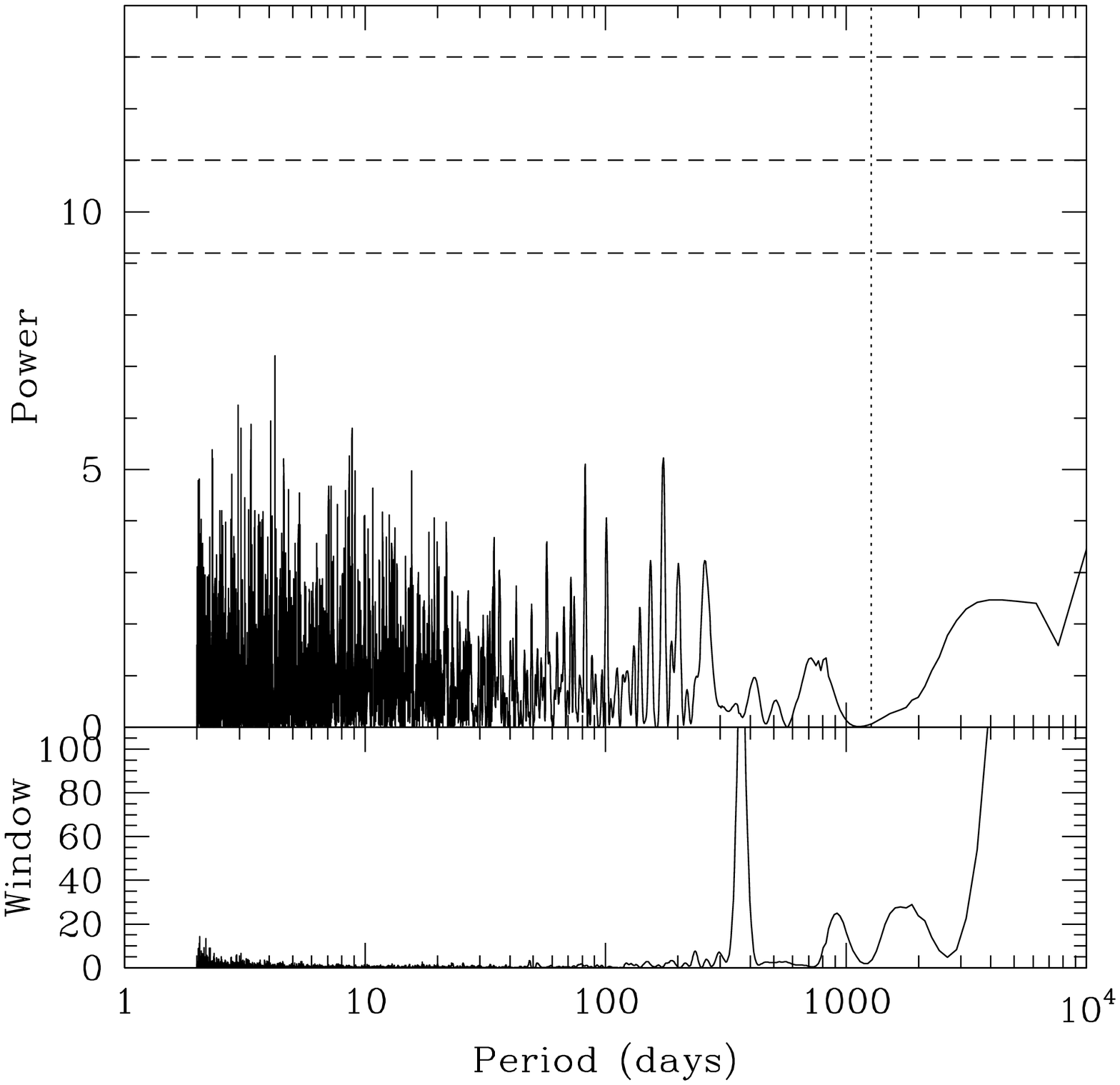}
\caption{Generalised Lomb-Scargle periodogram of ASAS photometry for 
HD\,86950.  A total of 515 epochs spanning 8.9 years yield no 
significant periodicities.  The 1291-day period of the planet is marked 
with a dotted line.  10\%, 1\%, and 0.1\% bootstrap-derived false alarm 
probabilities are shown as horizontal dashed lines. }
\label{86950pgram}
\end{figure}

For HD\,86950, we find a clear signal with $P=1270\pm$57 days and 
$K=49\pm$12\,\ms, corresponding to a planet with m~sin~$i$ of 
3.6$\pm$0.7\,\Mjup\ adopting a host star mass of 1.66\,\Msun.  The data 
and model fit are shown in Figure~\ref{86950fit}.  Examination of 8.9 
years (515 epochs) of ASAS photometry shows no periodicities of 
significance near the planet's orbital period (Figure~\ref{86950pgram}).  
The ASAS $V$ band photometry has a mean value of 7.47$\pm$0.02 
mag.

\begin{figure}
\includegraphics[scale=0.6, angle=270]{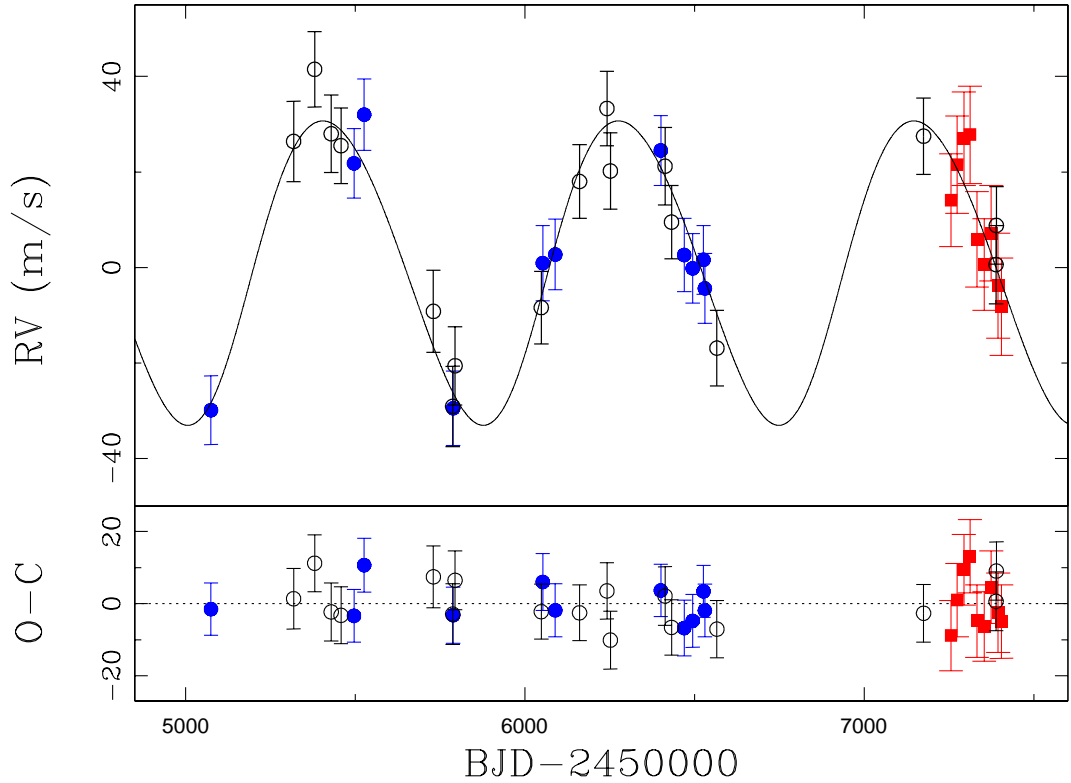}
\caption{Data and Keplerian fit for a 1.6\Mjup\ planet orbiting 
HD\,222076 (AAT -- blue, FEROS -- open, CHIRON -- red).  Error bars 
include 7\,\ms\ of jitter added in quadrature.  The rms about this fit 
is 5.9\,\ms. }
\label{222076fit}
\end{figure}

\begin{figure}
\includegraphics[scale=0.4]{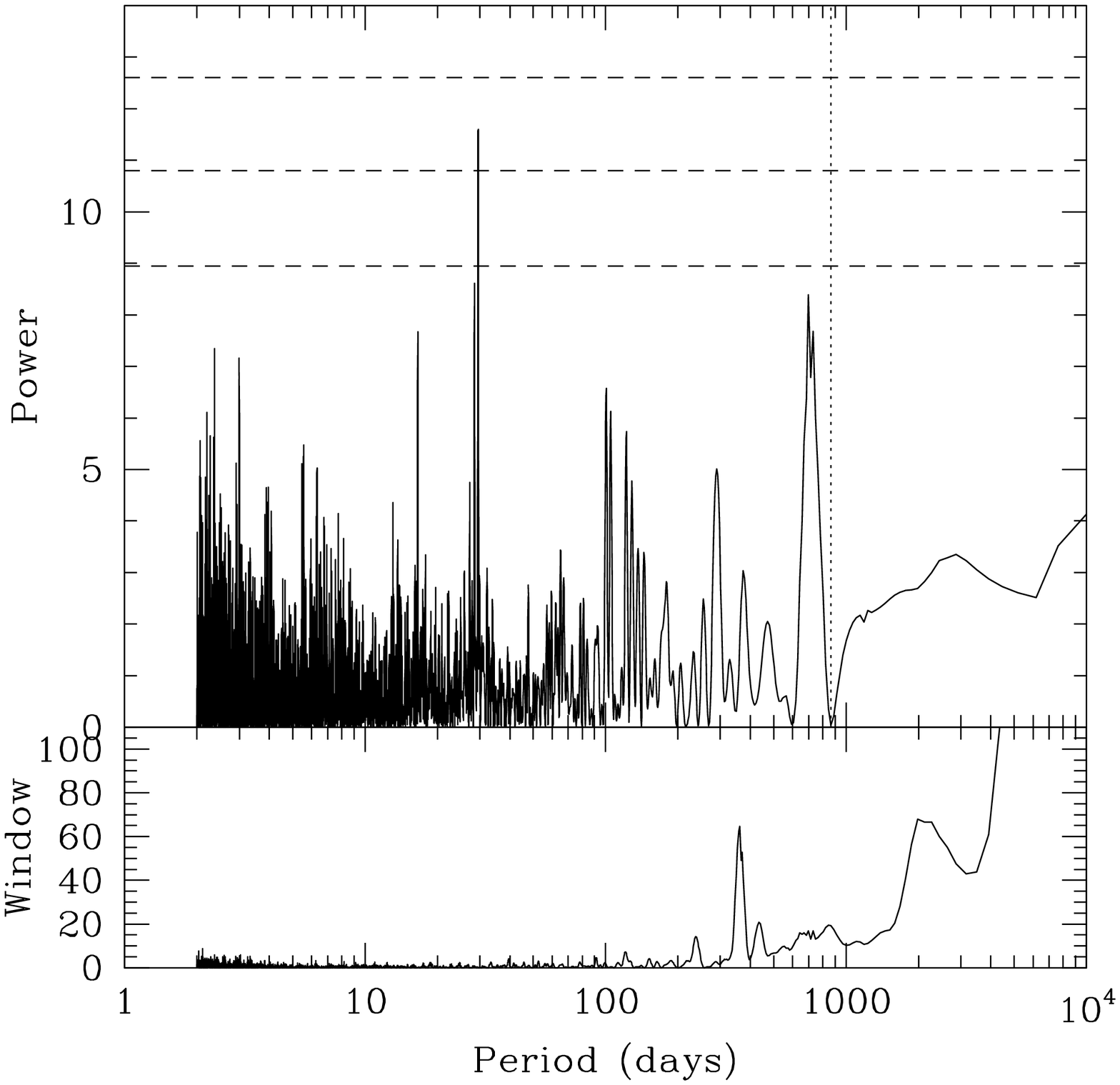}
\caption{Generalised Lomb-Scargle periodogram of ASAS photometry for 
HD\,222076.  A total of 425 epochs spanning 8.8 years yield no 
significant periodicities.  The 868-day period of the planet is marked 
with a dotted line.  10\%, 1\%, and 0.1\% bootstrap-derived false alarm 
probabilities are shown as horizontal dashed lines. }
\label{222076pgram}
\end{figure}

For HD\,222076, all three instruments clearly reveal a planet with 
$P=871\pm$19 days and m~sin~$i=1.56\pm$0.11\,\Mjup\ assuming a host star 
mass of 1.07\,\Msun\ \citep{ppps5}.  The orbit is nearly circular, and 
this fit has residuals of 5.9\,\ms\ (Figure~\ref{222076fit}).  Again, 
the ASAS photometry reveals no significant periodicities near the 
planet's orbital period (Figure~\ref{222076pgram}).  The ASAS 
$V$ band photometry has a mean value of 7.46$\pm$0.02 mag.  For this 
star, we have 5.67 years of FEROS spectra, uncontaminated by iodine, 
from which we can derive several activity indices (bisector velocity 
span, CCF FWHM, and S-index).  None of these indicators 
(Table~\ref{ferosact}) show any periodicities or correlations with the 
radial velocities.  We are thus confident that the observed velocity 
variations are due to an orbiting planet.

\subsection{H-alpha activity index}

In light of recent debate over the detection of planet-induced stellar 
reflex motion amidst the confounding effect of stellar activity 
\citep[e.g.][]{robertson14, anglada15, fischer16}, we discuss in this 
section the H$\alpha$ activity indices for the three stars considered 
here.  

Stellar activity can induce spurious velocity signatures that mimic the 
velocities produced by exoplanets.  This can appear at the stellar 
rotation period or its harmonics \citep{Boisse2011}.  Magnetic cycles 
can also produce radial velocity signals comparable in amplitude to 
orbiting planets \citep[e.g.][]{santos10, dum11, robertson15, faria16}.  
Multiple mechanisms can produce these line profile variations.  For 
example, variable levels of chromospheric activity can produce changes 
in the level of line profile reversal in some line cores, resulting in 
changes to both the line centroid and hence the measured radial velocity 
\citep{Martinez2010}.  These effects will also produce changes in the 
line's equivalent width (EW), and so measurement of the EW can provide 
an indicator of the presence of activity-induced radial velocity 
variations \citep{robertson14}.  Stellar activity is also seen to be 
correlated with the presence of starspots \citep{Berdyugina2005, 
Strassmeier2009, Davenport2015} and suppression of convection 
\citep{Haywood2016}, which will produce changes in line profile shapes 
and so line centroids, resulting in velocity jitter \citep{Reiners2010, 
Andersen2015}.  So EW measurements for a line-profile sensitive feature 
would be expected to be correlated with this source of jitter as well.

We have therefore measured the equivalent width of H$\alpha$ absorption 
as an indicator of variability in chromospheric activity.  Our analysis 
builds on that presented by \citep{robertson14}, with the addition of an 
automated algorithm for continuum normalisation and telluric 
contamination identification in the region of the H$\alpha$ line.  This 
method (detailed in the Appendix) has the advantage of being robust for 
arbitrary, slowly varying continua selection, without being parametric 
for the specific shape of the continuum or the location of specific 
absorption lines.

\begin{figure}
\plottwo{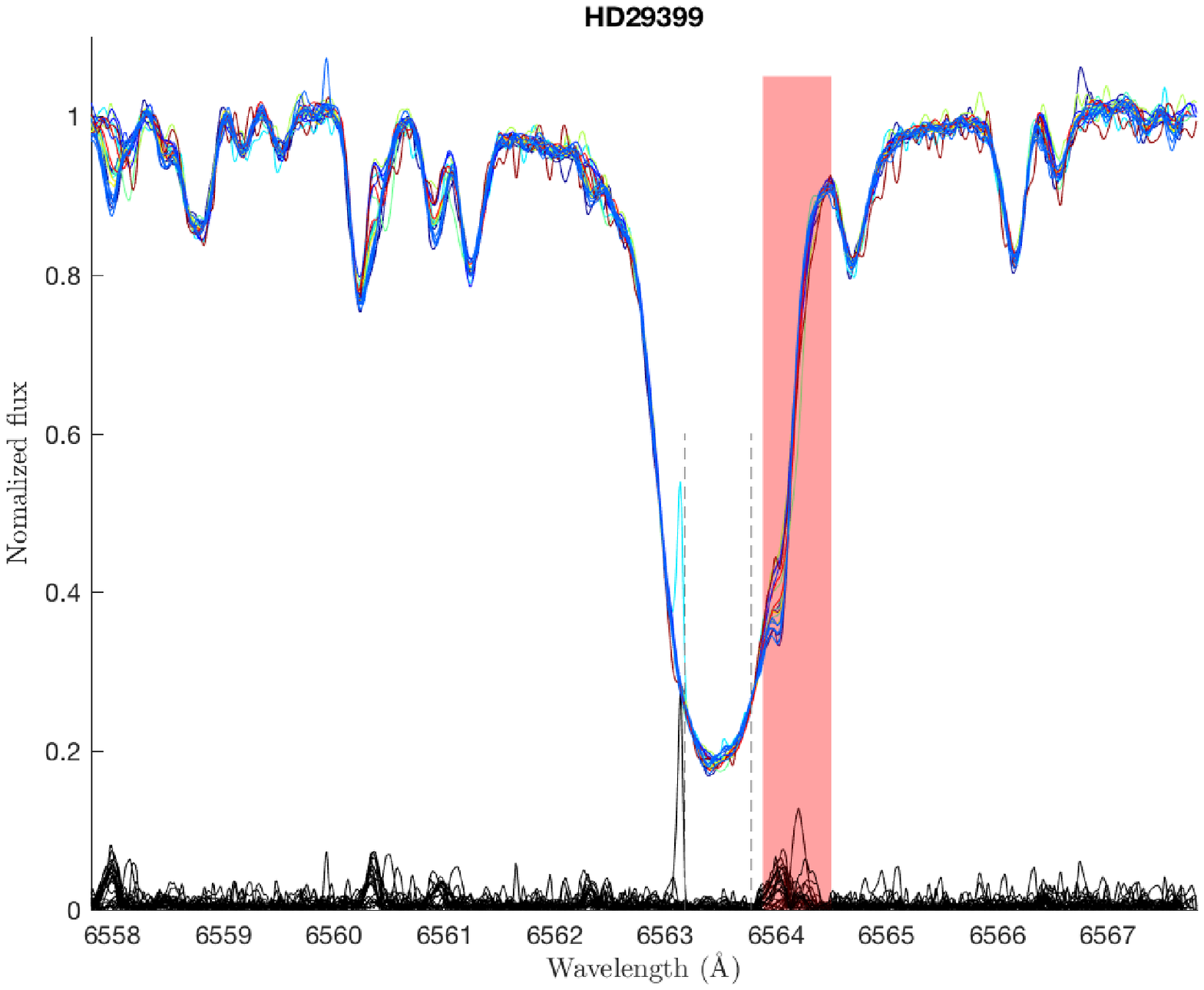}{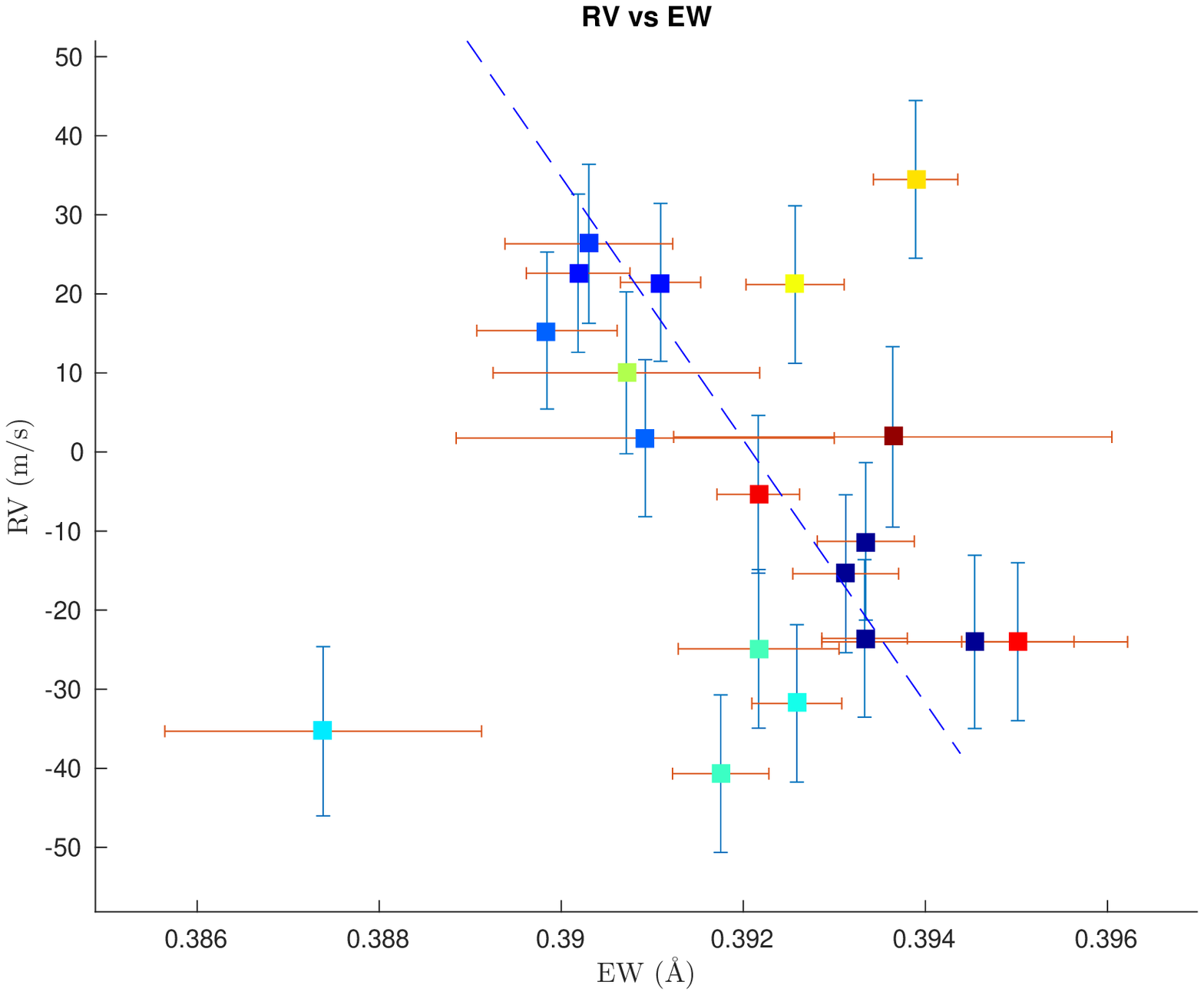}
\caption{Left: the individual normalized spectra stacked on top of each 
other.  The $0.5\mbox{\AA}$ H$\alpha$ region is labelled within the black 
dashed lines, whereas the telluric region is the shaded area within red 
dashed lines.  Differences from the template are shown at the bottom. 
Large residuals are due to telluric contamination.  Right: Radial velocity 
versus H$\alpha$ equivalent width.  The same epoch is presented in 
identical colours across these two panels, and the closeness in colour 
within the same panel represents the closeness in BJD. }
\label{figHD29399}
\end{figure}

\begin{figure}
\plottwo{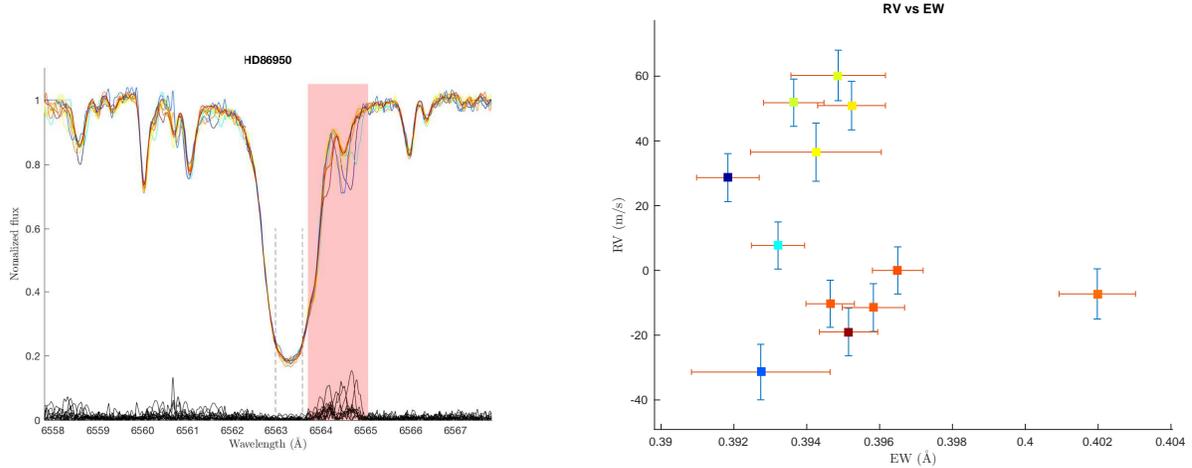}{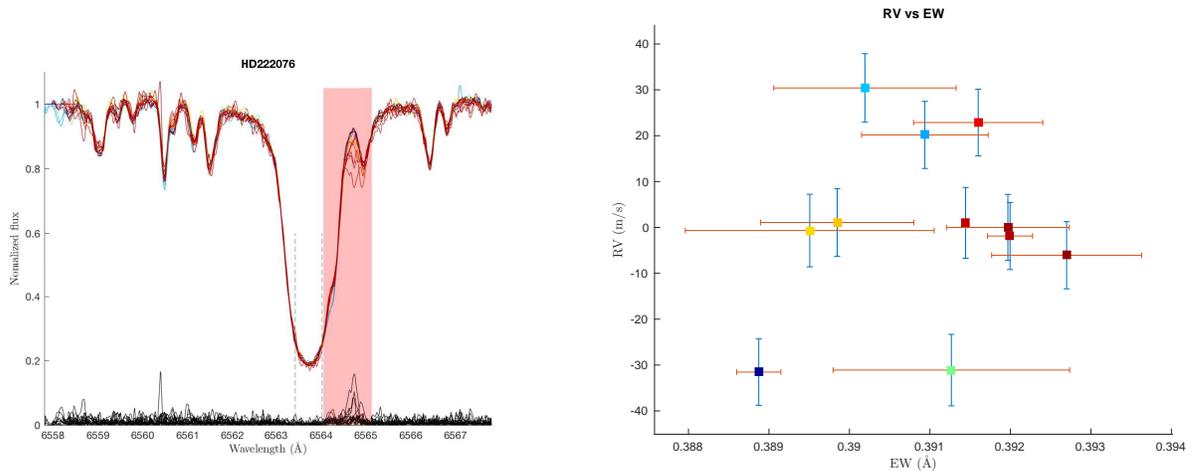}
\caption{Same as Fig.~\ref{figHD86950}, but for HD\,222076.}
        \label{figHD86950}
\end{figure}

\begin{figure}
\plottwo{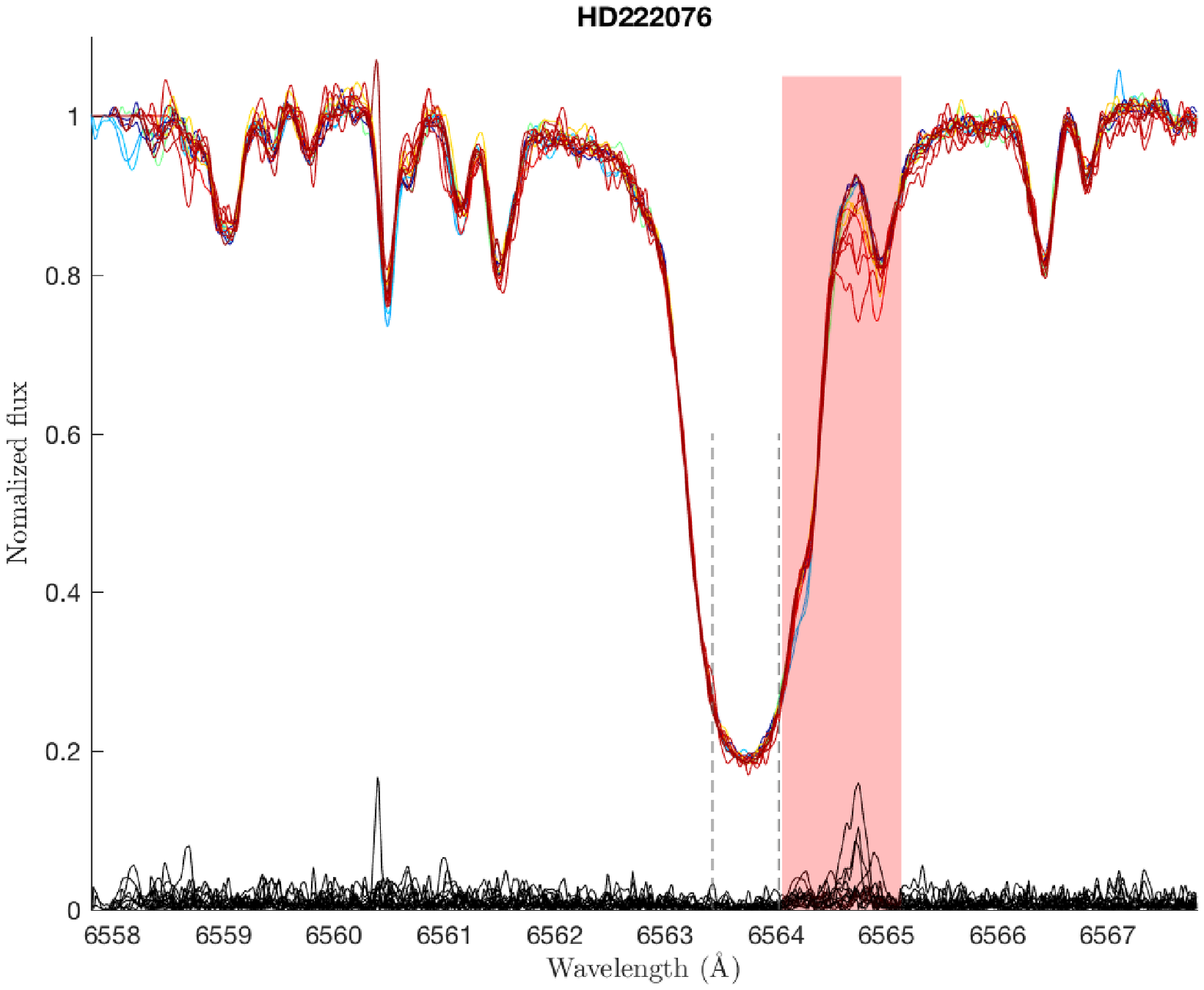}{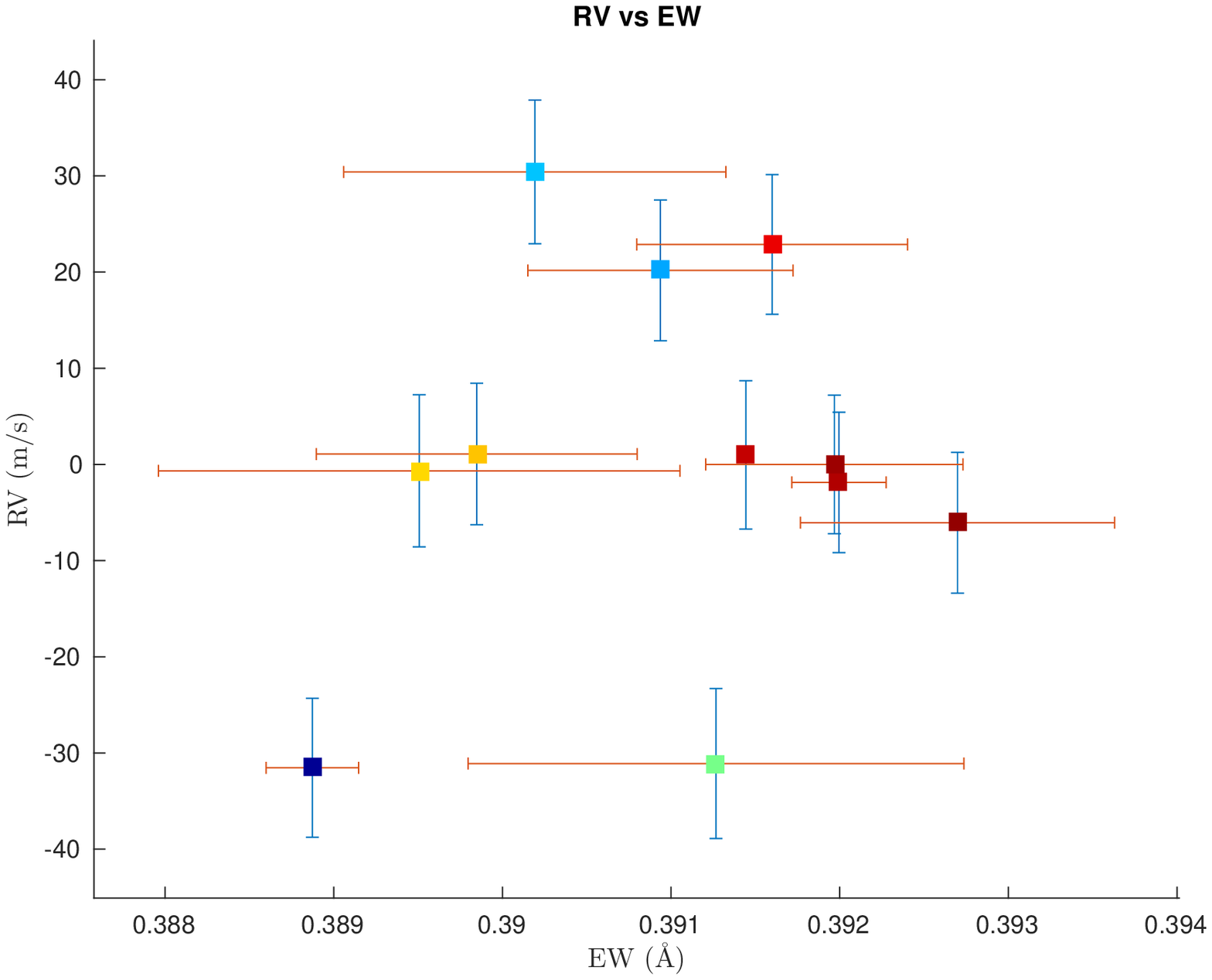}
\caption{Same as Fig.~\ref{figHD86950}, but for HD\,222076.}
        \label{figHD222076}
\end{figure}

We show the stacked spectra of the three stars featured in this work, 
along with their RV-EW relations (Fig.~\ref{figHD29399}-- 
\ref{figHD222076}).  For HD\,29399, where a strong $\sim$2-year 
periodicity was evident in the photometry and the radial velocities, we 
find a correlation between the H$\alpha$ equivalent width and the radial 
velocity, further evidence that the RV periodicity is intrinsic to the 
star and is most likely the result of a magnetic activity cycle 
as noted in Section 3.1.  For the candidate planet hosts HD\,86950 and 
HD\,222076, no correlations are seen from the RV-EW plots, supporting 
our claim that the detected RV variations are not due to chromospheric 
stellar activity.


\section{Discussion and Conclusions}

\begin{figure}
\includegraphics[scale=0.4]{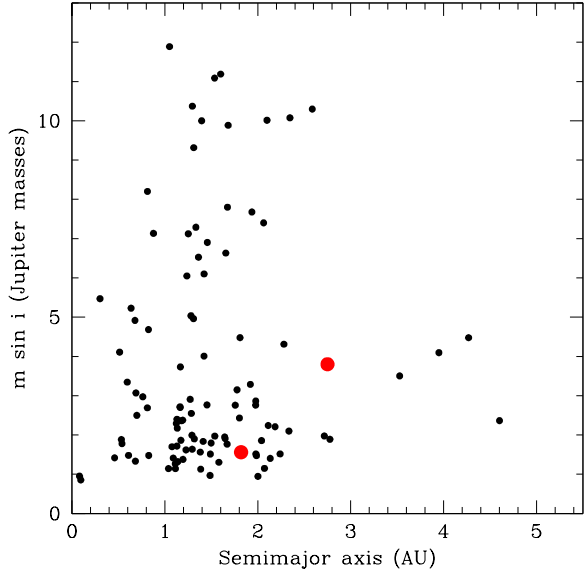}
\caption{Masses and semimajor axes of 107 planets orbiting giant stars 
(planet data obtained from the Exoplanet Orbit Database at 
exoplanets.org). The two larger red points are HD\,86950b and 
HD\,222076b.  These new planets are entirely typical of the evolved-host 
population. }
\label{comp}
\end{figure}

HD\,86950b and HD\,222076b are typical of the population of planets 
being found to orbit evolved stars, which are generally beyond 
1\,AU and with masses greater than 1\,Mjup\ \citep[e.g.][]{lovis07, 
dollinger09, bowler10, jones14}.  Figure~\ref{comp} places these two 
planets in context with the other planets known to orbit giant stars 
(i.e.~stars with log $g<3.5$).  Of the 107 such planets 
confirmed\footnote{\url{http://exoplanets.org}, accessed 2016 July}, the 
median semimajor axis is 1.38\,AU, and the median m~sin~$i$ is 
1.38\,\Mjup.

\subsection{Possible debris discs}

HD\,86950 has been identified by \cite{Macdonald2012} as having a possible 
infrared excess based on the presence of excess emission at $9\mu$m in 
the AKARI/IRC All-Sky Survey \citep{Ishihara2010}, with a fractional 
luminosity ($L_{dust}/L_{star}$) of $\sim 1.2\times\,10^{-3}$.  The 
presence of both a planetary system \textit{and} a debris disc around 
HD\,86950 would make it a nearly-unique object amongst sub-giant stars, 
joining $\kappa$ CrB (HD\,142091) as one of very few examples of an 
evolved star hosting both a debris disc and exoplanet \cite{Bonsor2013}.  
Confirmation of this excess (and a better determination of the disc's 
properties if confirmed) is therefore critical.

\begin{figure}
\plottwo{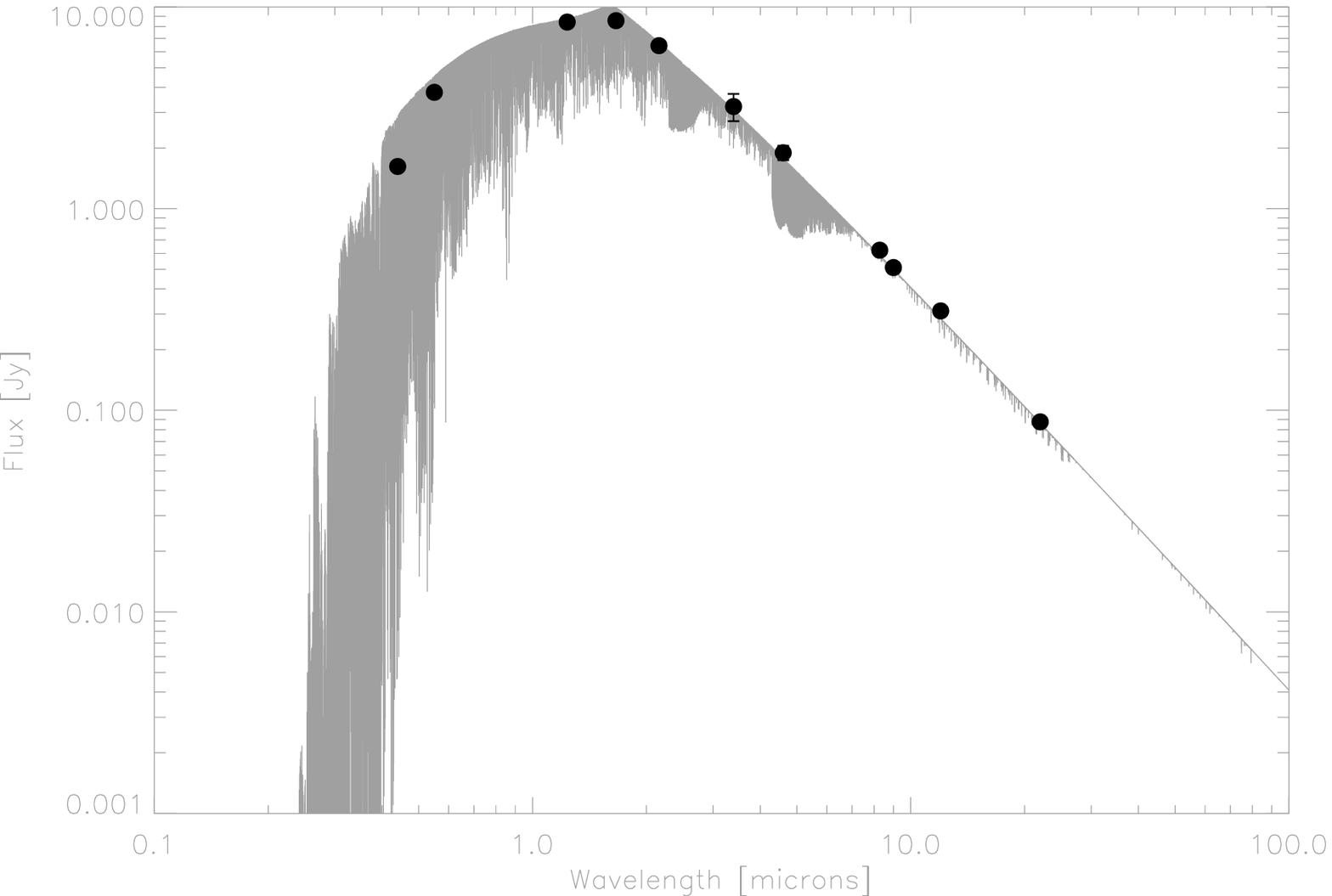}{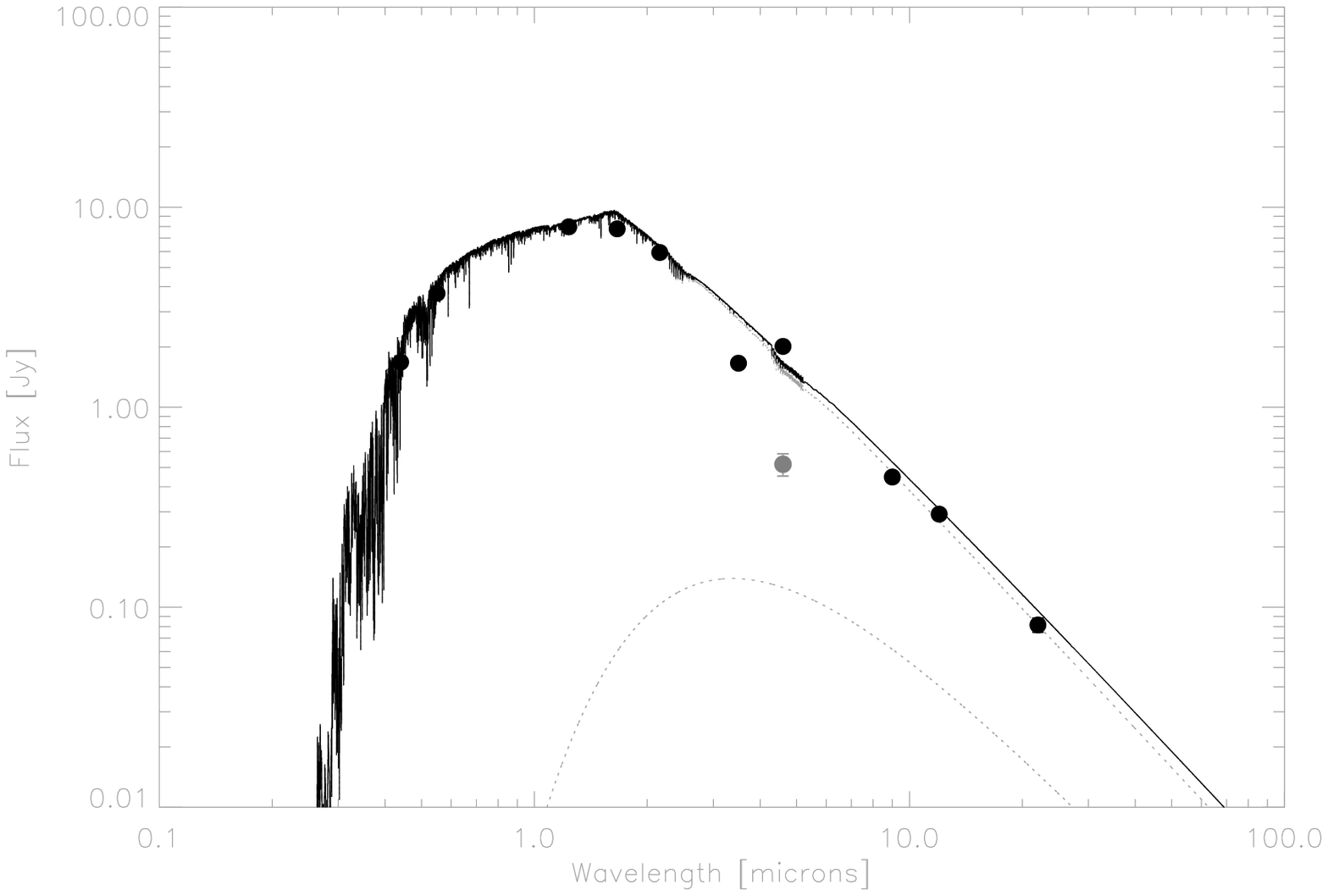}
\caption{Left: Spectral energy distribution of HD\,86950.  The 
photometric data compiled from literature sources are shown in black, 
with 1-$\sigma$ uncertainties.  The stellar photosphere model is shown 
in grey, and has been scaled according to the assumed stellar radius and 
parallax-derived distance (i.e. it is not a least-squares fit to the 
photometry).  No evidence of infrared excess at the observed wavelengths 
is present.  Right: Same, but for HD\,222076. }
\label{hd86950_sed}
\end{figure}

We compiled a spectral energy distribution from photometry spanning 
optical to mid-infrared wavelengths, including optical $BV$, 
near-infrared 2MASS $JHK_{s}$ \citep{Skrutskie2006}, and mid-infrared 
MSX \citep{Egan2003} and WISE \citep{Wright2010} measurements, in 
addition to the AKARI 9~$\mu$m datum.  We fit the stellar photospheric 
emission with a model from the BT-SETTL/Nextgen stellar atmospheres grid 
appropriate for the spectral type (K0III; $T_{\rm eff} = 4750~$K, log 
$g$ = 2.0, [Fe/H] = 0.0), and scaled to the stellar radius and distance 
\citep{Wittenmyer2015,vl07}.  We colour corrected the AKARI and WISE 
flux densities assuming blackbody emission from the star.  The resulting 
spectral energy distribution is shown in the left panel of 
Figure~\ref{hd86950_sed}.  No evidence of significant excess (i.e. 
($F_{\rm obs} - F_{\star} / \sigma_{\rm obs}) > 3$) from the system is 
observed out to wavelengths $\sim$~22~$\mu$m, ruling out the warm, 
bright disc inferred from \citet{Macdonald2012}.  The previous 
identification of infrared excess from HD\,86950 can be attributed to 
the AKARI flux density not being colour corrected.

We performed a similar analysis of the available photometric data for 
HD\,222076.  We find no evidence for an excess, with the maximum at 
9$\mu$m and 1.8$\sigma$ (Figure~\ref{hd86950_sed}, right panel).  
\citet{Macdonald2012} likewise find no excess in their analysis.

\begin{figure}
\plotone{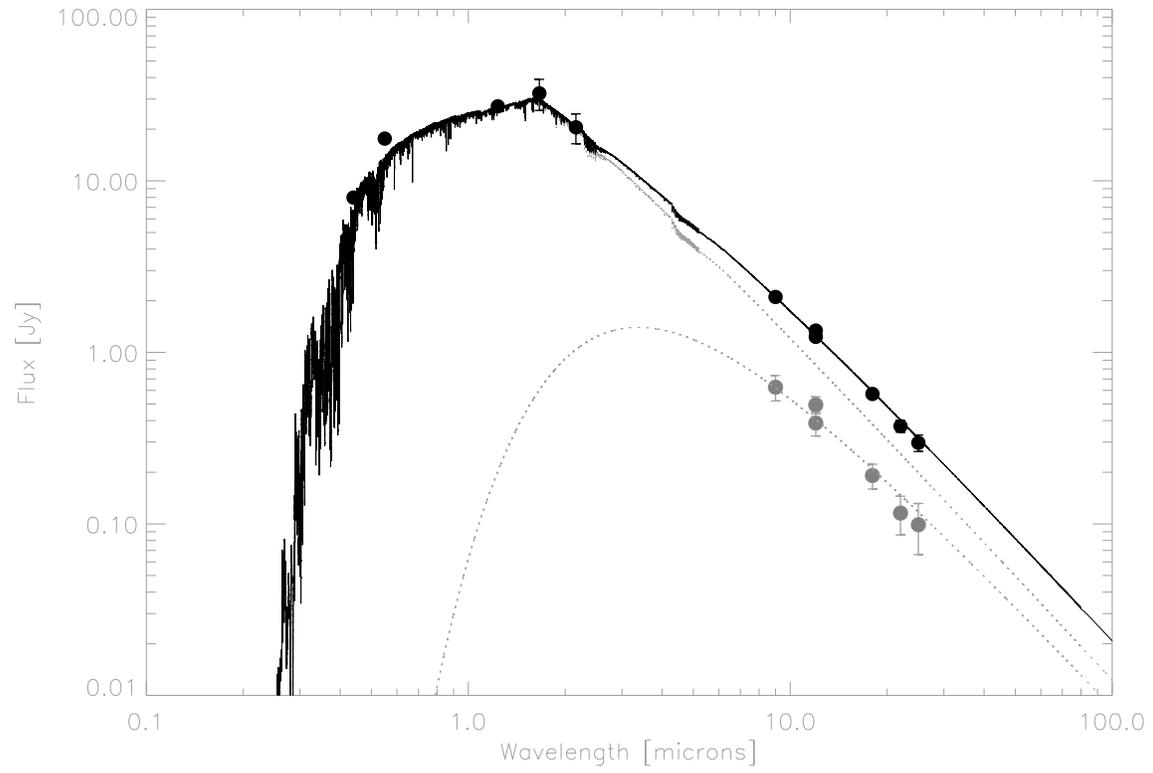}
\caption{Spectral energy distribution of HD\,29399.  An excess is 
evident at the 3-8$\sigma$ level, and can be fitted with a star+dust 
model using 1500\,K dust (lower dotted curve). }
\label{hd29399_sed}
\end{figure}

While we find that HD\,29399 does not host a planet, we note that 
\citet{Macdonald2012} also identify it as having an excess, with 
($L_{dust}/L_{star}$) $\sim 3.6\times\,10^{-3}$.  Using the same 
approach as described above, we show the spectral energy distribution 
for HD\,29399 in Figure~\ref{hd29399_sed}.  Photometry for the 
targets was compiled from Johnson BV, 2MASS JHKs \citep{cutri2003}, WISE 
All-Sky Survey \citep{Wright2010}, and the Akari IRC All-Sky Survey 
\citep{Ishihara2010}.  We avoided using WISE W1 and W2 photometry due to 
known saturation issues for bright ($V<8$) stars.  We fit the stellar 
photospheric emission with the following model atmosphere consistent 
with the physical parameters given in Table~\ref{stellarparams}: $T_{\rm 
eff} = 4800~$K, log $g$ = 3.5, [Fe/H] = 0.0.  Adopting a stellar radius 
of 3.97\Rsun\ and the distance as given in 
Table~\ref{stellarparams}, we find an infrared excess at the 3-8$\sigma$ 
level.  This can be fitted with a star+dust model using 1500K dust 
($\chi^{2}_{\nu}=1.085$).  The star is not yet so evolved (nor is its 
photospheric temperature cool enough) that it would be expected to be 
surrounded by a dusty envelope, one potential origin of the excess. 
Likewise, the radial velocities rule out the presence of a cool binary 
stellar companion.  If we adopt a 25\% larger stellar radius of 
$\sim$5\Rsun, then the excess disappears and would be consistent with 
the poorly constrained near-infrared photometry for this star.  However, 
assuming we are confident in the derivation of our stellar parameters 
and that the circumstellar dust origin for the excess holds, such hot 
dust might therefore be produced by e.g. delivery of comets from an 
outer, cool debris belt.  Additional measurements of the spectral energy 
distribution at wavelengths $> 30\,\mu$ to search for such a debris belt 
would be of value in this case.

\subsection{The giant planet-metallicity relation}

\begin{figure}
\plotone{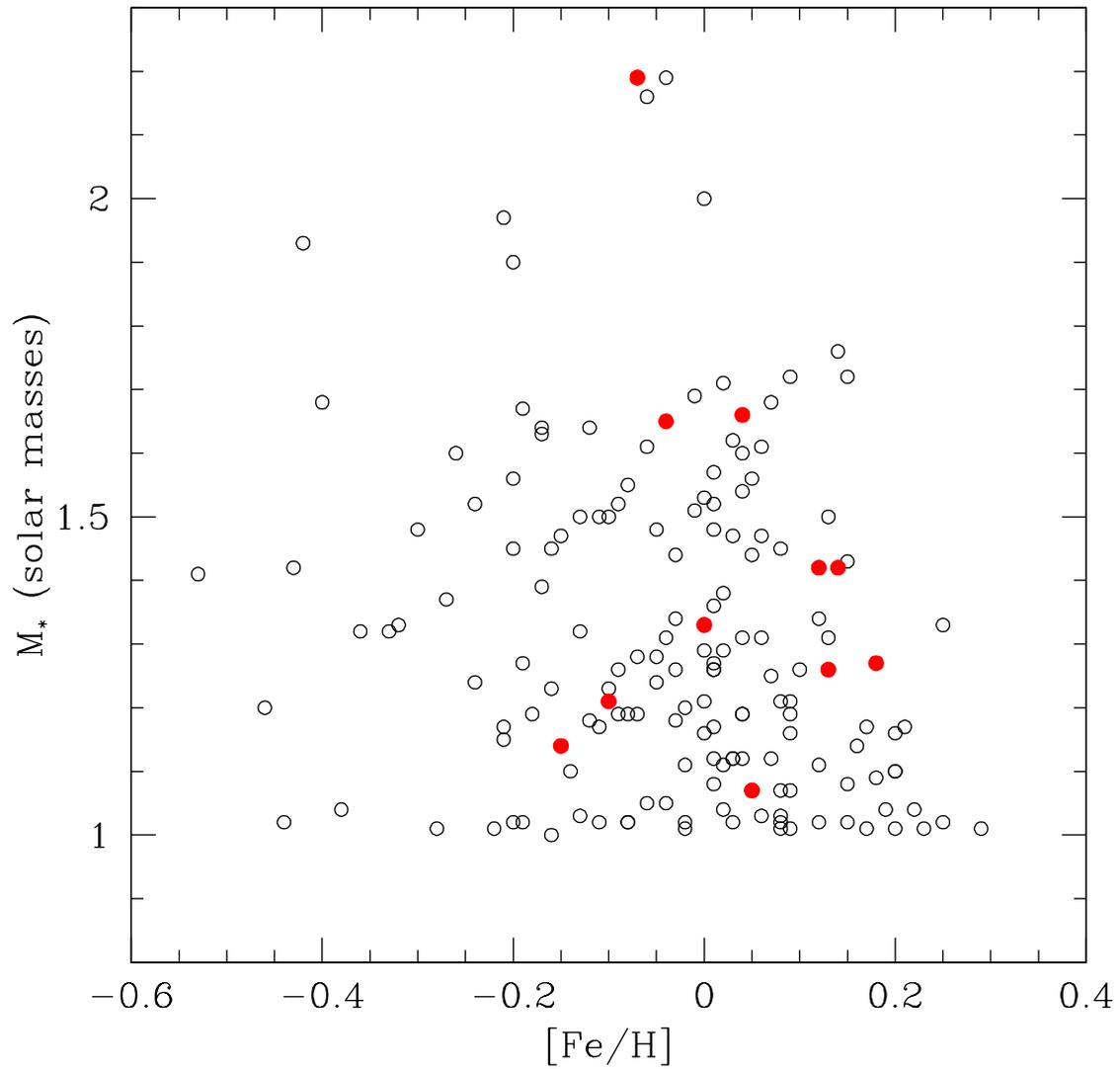}
\caption{Host-star mass versus metallicity for 164 evolved stars in the 
PPPS.  Stellar parameter data have been taken from \citet{ppps5}.  
Planet hosts are shown as filled red circles. }
\label{massvsfeh}
\end{figure}

Following recent work by \citet{reffert15} and \citet{jones16}, we now 
examine the results of the PPPS in light of the planet-metallicity 
relation for evolved hosts as noted by those authors.  Overall, our PPPS 
sample of 164 stars has 11 confirmed planet hosts, comparable to the 10 
planet hosts in 166 stars reported by the EXPRESS survey 
\citep{jones16}.  \citet{reffert15} reported 15 secure planet hosts (and 
a further 20 candidates) among 373 giant stars in their Lick Observatory 
sample.  Figure~\ref{massvsfeh} shows the host-star mass versus 
metallicity [Fe/H] for 164 PPPS stars; the 11 confirmed hosts are shown 
as red points.

For ease of comparison, we cast our results into the same stellar 
mass-metallicity bins as used by \citet{reffert15}.  Table~\ref{metal} 
shows the PPPS results in these bins; \citet{jones16} presented the 
EXPRESS results in the same format in their Table 3.  Our results are 
broadly consistent; a key difference is that the PPPS sample has very 
few targets of super-solar metallicity compared to the EXPRESS and Lick 
surveys.  This results in the highest-metallicity bins being sparsely 
populated and not quite amenable to a full analysis including the 
dimension of host-star mass (as was performed for the 373-star sample 
presented in Reffert et al.~2015).  

\begin{figure}
\plotone{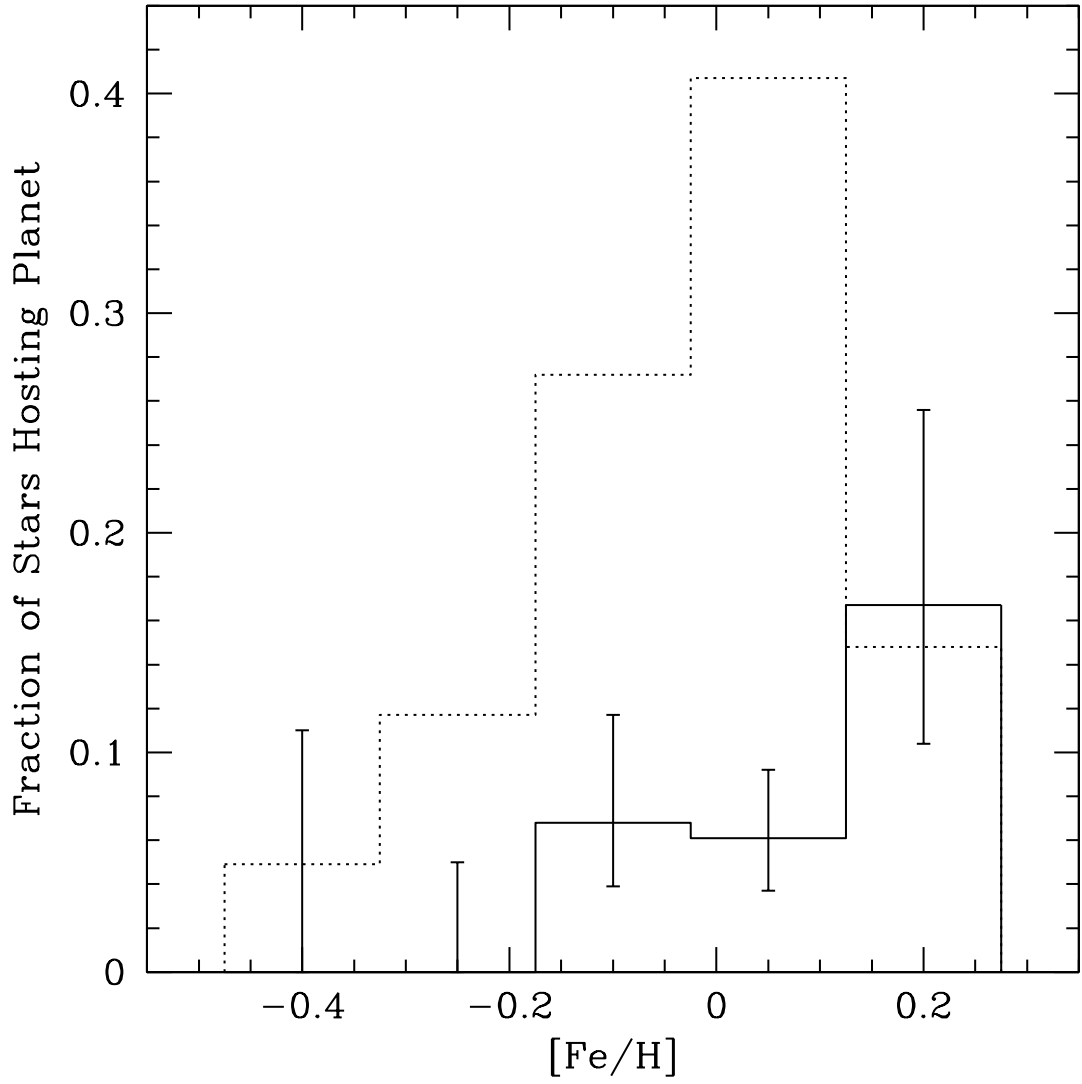}
\caption{Occurrence rate as a function of metallicity for PPPS targets 
with confirmed planets.  The dashed histogram shows the parent sample 
distribution.  The bins are the same as defined in Figure 8 of 
\citet{jones16}. }
\label{freq_feh}
\end{figure}

We next investigate the occurrence rate only as a function of [Fe/H] as 
shown in Figure~\ref{freq_feh}.  For ease of comparison with 
\citet{jones16}, we use the same bins as from their Figure 8.  The 
68.3\% binomial confidence intervals are computed after 
\citet{wilson1927}, which is noted by \citet{brown01} as a preferable 
method for small sample sizes.

Consistent with \citet{hekker07}, \citet{reffert15}, and 
\citet{jones16}, we also show an increasing planet occurrence rate with 
stellar metallicity.  The two samples, of planet hosts and the parent 
sample, are shown by the K-S test to be drawn from the same distribution 
with a K-S probability of $P=0.708$, i.e.~a 70.8\% probability that they 
are from the same underlying distribution. 

\begin{figure}
\plottwo{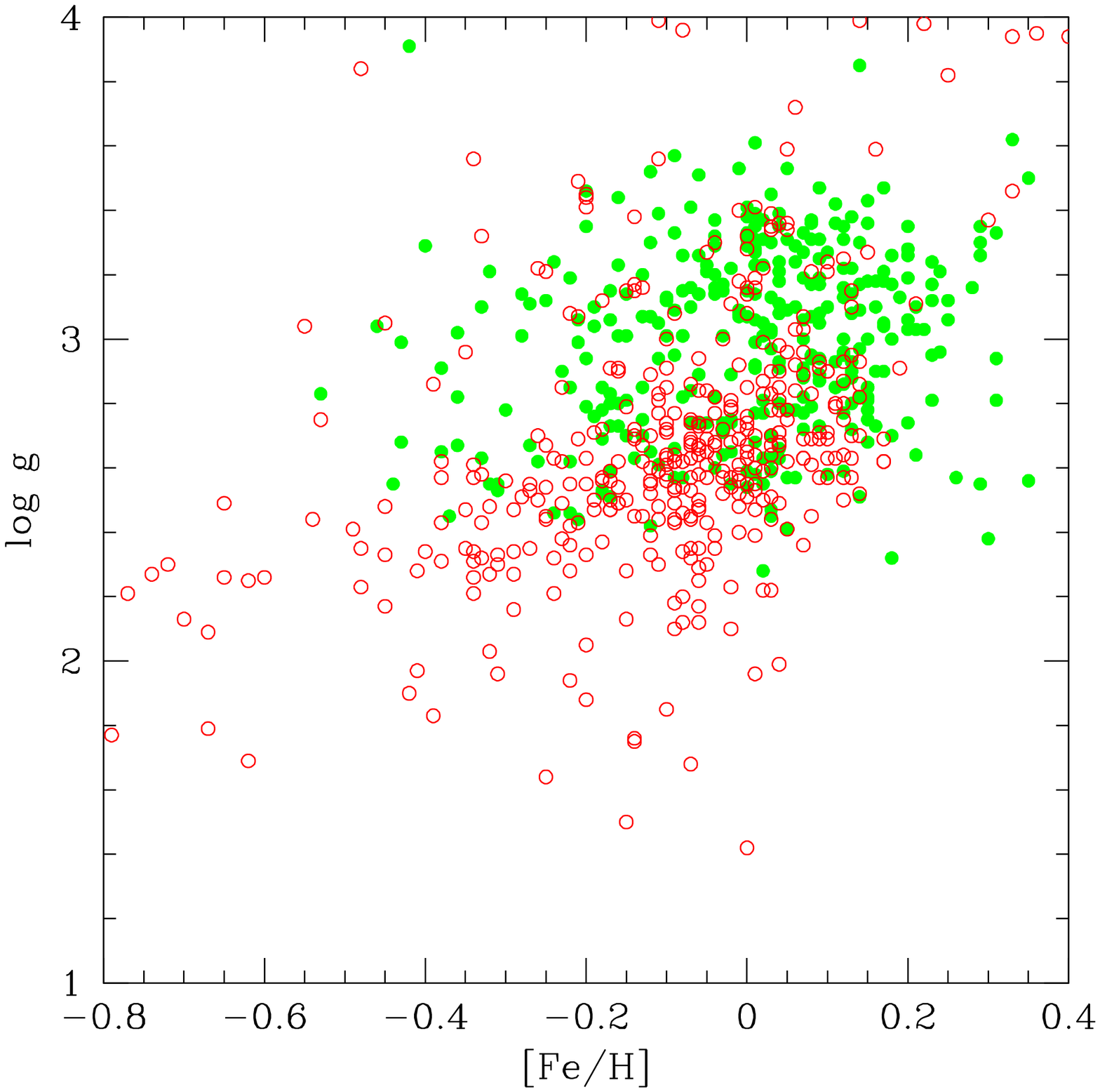}{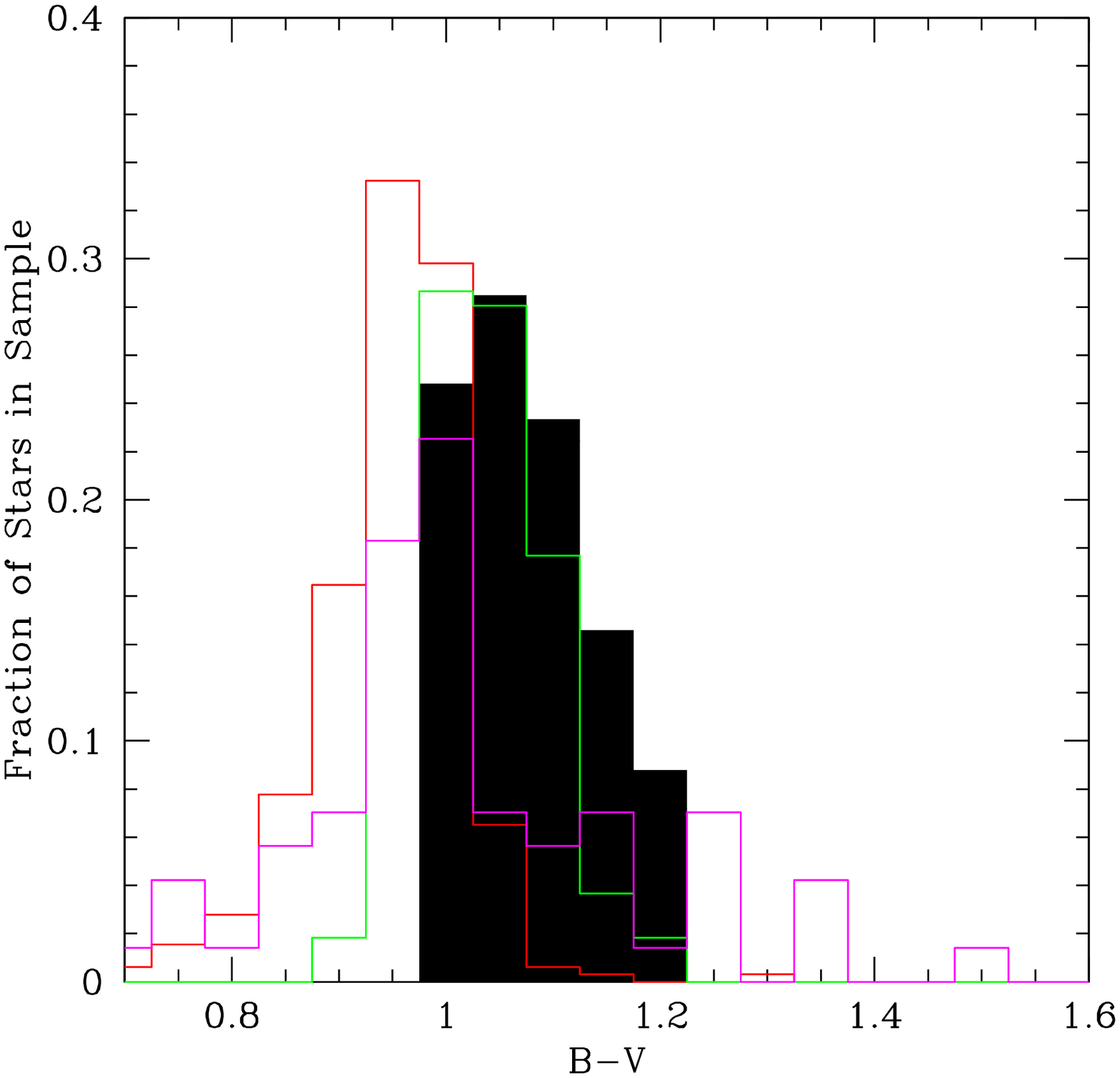}
\caption{Left: Surface gravity versus [Fe/H] for samples of evolved stars 
for which a planet-metallicity correlation is evident (green filled 
circles) and not evident (red open circles).  The former are drawn from 
\citet{jones16} and \citet{ppps5}, while the latter are drawn from 
\citet{takeda08} and \citet{mortier13}.  Right: Normalized distribution of 
colours of evolved stars from four surveys: solid -- \citet{ppps5}; green 
-- \citet{jones16}; red -- \citet{takeda08}; magenta -- \citet{mortier13}.  
Surveys not finding a planet-metallicity correlation have largely excluded 
stars redward of $(B-V)=1.0$, a bias proposed by \citet{mortier13} as a 
possible source of the discrepant conclusions reached by these surveys. }
\label{colours}
\end{figure}

Our findings of a positive correlation between metallicity and 
planet occurrence are in contrast to the results of some other evolved 
star surveys \citep[e.g.][]{pasquini07, takeda08, mortier13}.  The 
disagreement most likely arises from different selection criteria for the 
samples of evolved stars observed by the various groups.  
\citet{mortier13} pointed out that since most surveys choose a colour 
cutoff $(B-V)\leq$1.0, the more metal-rich, low-gravity stars are 
excluded.  To illustrate this, Figure~\ref{colours} plots the distribution 
of log\,$g$ versus [Fe/H] for the targets of four surveys, two of which 
find a planet-metallicity correlation \citep{jones16, ppps5}, and two of 
which do not \citep{takeda08, mortier13}.  The surveys which did not find 
a correlation are shown as red open circles, generally lying at lower 
[Fe/H] and lower surface gravity than the others.  \citet{mortier13} 
showed a similar result in their Figure~6.  We also show the differences 
in (B-V) colours from these four surveys in the right panel of 
Figure~\ref{colours}.  Again, the surveys finding a planet-metallicity 
correlation tend to sample redder stars with $(B-V)\ge$1.0. 

These results from the PPPS do not yet account for survey incompleteness 
induced by nonuniform detectability \citep[e.g.][]{howard10, etaearth, 
newjupiters}.  A more comprehensive analysis of the overall survey 
completeness is the subject of a forthcoming paper.


\acknowledgements

We gratefully acknowledge the efforts of PPPS guest observers Brad 
Carter, Hugh Jones, and Simon O'Toole.  This research has made use of 
NASA's Astrophysics Data System (ADS), and the SIMBAD database, operated 
at CDS, Strasbourg, France.  This research has also made use of the 
Exoplanet Orbit Database and the Exoplanet Data Explorer at 
exoplanets.org \citep{wright11, han14}.

\software{MOOG (Sneden 1973), Kurucz ATLAS9 (Castelli \& Kurucz 2003)}



\clearpage

\begin{deluxetable}{lrrrr}
\tabletypesize{\scriptsize}
\tablecolumns{5}
\tablewidth{0pt}
\tablecaption{Summary of observations}
\tablehead{
\colhead{Star} & \colhead{$N_{obs}$} & \colhead{Span (days)} & 
\colhead{Mean uncertainty (\ms)} & \colhead{Instrument} }
\startdata
\label{obslog}
HD 29399   & 22 & 2373 & 1.8 & UCLES \\
           & 7 & 79 & 4.2 & CHIRON \\ 
HD 86950   & 13 & 1879 & 2.8 & UCLES \\
           & 7 & 66 & 3.9 & CHIRON \\
HD 222076  & 11 & 1456 & 2.5 & UCLES \\
        & 17 & 2072 & 4.3 & FEROS \\
        & 9  & 149  & 7.2 & CHIRON \\
\enddata
\end{deluxetable}


\begin{deluxetable}{lrrr}
\tabletypesize{\scriptsize}
\tablecolumns{4}
\tablewidth{0pt}
\tablecaption{Radial velocities for HD\,29399}
\tablehead{
\colhead{BJD-2400000} & \colhead{Velocity (\ms)} & \colhead{Uncertainty
(\ms)} & \colhead{Instrument}}
\startdata
\label{29399vels}
54866.93634  &    -24.01  &    4.68  & AAT \\
54867.96653  &    -23.57  &    1.04  & AAT \\
54869.89726  &    -15.40  &    1.32  & AAT \\
54870.90323  &    -11.30  &    1.15  & AAT \\
55074.31485  &     22.61  &    1.40  & AAT \\
55075.31193  &     21.45  &    1.23  & AAT \\
55140.16735  &     26.33  &    1.76  & AAT \\
55225.01910  &      1.75  &    0.68  & AAT \\
55225.91346  &     15.36  &    0.77  & AAT  \\
55226.92603  &      6.46  &    0.91  & AAT \\
55455.23879  &    -35.32  &    4.05  & AAT \\
55525.12752  &    -31.79  &    1.00  & AAT \\
55580.91947  &    -40.68  &    1.07  & AAT  \\
55601.91656  &    -24.89  &    1.62  & AAT \\
55783.26560  &     10.02  &    2.60  & AAT  \\
55879.16028  &     12.69  &    1.31  & AAT \\
55906.05799  &     21.17  &    1.06  & AAT \\
55971.05135  &     34.47  &    1.21  & AAT \\
56344.94543  &    -23.99  &    1.32  & AAT  \\
56375.87846  &     -5.35  &    1.24  & AAT \\
56526.22058  &      1.91  &    5.67  & AAT \\
57238.28861  &     -5.13  &    1.55  & AAT  \\
\hline
57298.82760  &     -6.50  &    4.30  & CHIRON \\
57299.74140  &     -8.70  &    4.80  & CHIRON \\
57308.68060  &     19.90  &    4.30  & CHIRON  \\
57324.81200  &      9.20  &    4.40  & CHIRON  \\
57332.68480  &      2.30  &    3.90  & CHIRON  \\
57353.71510  &     -6.00  &    4.20  & CHIRON \\
57377.59490  &    -10.10  &    3.80  & CHIRON  \\
\enddata
\tablecomments{The velocities shown are relative to
instrument-specific zero points.}
\end{deluxetable}

\clearpage

\begin{deluxetable}{lrrr}
\tabletypesize{\scriptsize}
\tablecolumns{4}
\tablewidth{0pt}
\tablecaption{Radial velocities for HD\,86950}
\label{86950vels}
\tablehead{
\colhead{BJD-2400000} & \colhead{Velocity (\ms)} & \colhead{Uncertainty
(\ms)} & \colhead{Instrument}}
\startdata
54866.15767  &     28.67  &    2.43  & AAT \\
55252.08271  &    -31.39  &    4.95   & AAT \\
55381.85809  &    -42.55  &    1.85  & AAT \\
55581.16782  &      7.70  &    1.97  & AAT \\
55969.12369  &     51.80  &    2.08  & AAT \\
55994.08537  &     60.23  &    3.44  & AAT \\
56051.93221  &     36.53  &    5.64  & AAT \\
56090.89459  &     50.92  &    2.72  & AAT \\
56344.07002  &     -7.25  &    3.33  & AAT \\
56375.04442  &    -11.47  &    2.34  & AAT  \\
56377.01080  &    -10.30  &    1.98  & AAT \\
56399.95118  &      0.00  &    1.99  & AAT \\
56745.02214  &    -19.00  &    2.18  & AAT \\
\hline
57394.72380  &      3.40  &    4.10  & CHIRON  \\
57404.86600  &      4.70  &    3.80  & CHIRON  \\
57406.82630  &      4.90  &    4.20  & CHIRON \\
57407.85770  &     -7.40  &    4.50  & CHIRON \\
57445.62610  &      5.30  &    3.70  & CHIRON  \\
57445.64010  &     -0.20  &    3.20  & CHIRON \\
57460.60950  &    -10.60  &    3.80  & CHIRON \\
\enddata
\tablecomments{The velocities shown are relative to instrument-specific 
zero points, which are free parameters in the fitting process and are 
given in Table~\ref{planetparams}.}
\end{deluxetable}


\begin{deluxetable}{lrrr}
\tabletypesize{\scriptsize}
\tablecolumns{4}
\tablewidth{0pt}
\tablecaption{Radial velocities for HD\,222076}
\label{222076vels}
\tablehead{
\colhead{BJD-2400000} & \colhead{Velocity (\ms)} & \colhead{Uncertainty
(\ms)} & \colhead{Instrument}}
\startdata
55074.25071  &     -31.5  &    1.8  & AAT \\
55495.99749  &      20.2  &    2.1  & AAT  \\
55525.93596  &      30.4  &    2.6  & AAT  \\
55788.26403  &     -31.1  &    3.4  & AAT  \\
56052.29321  &      -0.7  &    3.7   & AAT \\
56089.28126  &       1.1  &    2.3  & AAT \\
56400.31160  &      22.9  &    2.0   & AAT \\
56469.30504  &       1.0  &    3.2  & AAT  \\
56494.25818  &      -1.9  &    2.1   & AAT  \\
56526.10530  &       0.0  &    1.7  & AAT  \\
56530.11951  &      -6.1  &    2.1   & AAT \\
\hline
55317.88970  &      16.0  &    4.6  & FEROS  \\
55379.90380  &      31.1  &    3.6  & FEROS  \\
55428.75550  &      17.6  &    4.0  & FEROS \\
55457.72790  &      15.1  &    3.6  & FEROS \\
55729.91020  &     -19.6  &    5.0  & FEROS  \\
55786.89910  &     -39.4  &    4.6  & FEROS \\
55793.83430  &     -31.0  &    4.2  & FEROS \\
56047.93430  &     -18.7  &    3.0  & FEROS \\
56160.79830  &       7.7  &    3.3  & FEROS \\
56241.62910  &      23.0  &    3.4  & FEROS \\
56251.66450  &       9.8  &    3.9  & FEROS  \\
56412.86280  &      10.8  &    4.1  & FEROS  \\
56431.84030  &      -0.9  &    3.1  & FEROS  \\
56565.68380  &     -27.2  &    3.7  & FEROS  \\
57174.87230  &      17.1  &    3.8  & FEROS \\
57388.55510  &      -9.8  &    4.2  & FEROS  \\
57389.55090  &      -1.6  &    4.0  & FEROS \\
\hline
 57255.75540  &       3.9  &    6.7  & CHIRON \\
 57273.74490  &      11.3  &    7.4  & CHIRON  \\
 57293.66430  &      16.7  &    6.8  & CHIRON  \\
 57311.66690  &      17.5  &    7.4  & CHIRON  \\
 57332.57240  &      -4.3  &    7.2  & CHIRON  \\
 57353.53290  &      -9.7  &    6.6  & CHIRON \\
 57374.53140  &      -3.0  &    7.2  & CHIRON  \\
 57394.53520  &     -14.0  &    8.5  & CHIRON \\
 57404.54130  &     -18.5  &    7.4  & CHIRON \\
\enddata
\tablecomments{The velocities shown are relative to instrument-specific
zero points, which are free parameters in the fitting process and are
given in Table~\ref{planetparams}.}
\end{deluxetable}

\clearpage

\begin{deluxetable}{lllllll}
\tabletypesize{\scriptsize}
\tablecolumns{7}
\tablewidth{0pt}
\tablecaption{Stellar Parameters }
\tablehead{
\colhead{} & \multicolumn{2}{c}{HD 29399} & \multicolumn{2}{c}{HD 86950} & 
\multicolumn{2}{c}{HD 222076} \\
\colhead{Parameter} & \colhead{Value} & \colhead{Ref.} & \colhead{Value} 
& \colhead{Ref.} & \colhead{Value} & \colhead{Ref.} }
\startdata
\label{stellarparams}
Spec.~Type         & K1 III            & 2                     & K1 III         & 8                      & K0 III          & 7                             \\
$(B-V)$            & 1.03            & 9                     &   1.09           &  10                    & 1.030           & 8                             \\
$E(B-V)$           & 0.0053            & 1                     & 0.0173        & 1                       & 0.0121          & 1                              \\
$A_V$              & 0.0165            & 1                     & 0.0537        & 1                       & 0.0374          & 1                              \\
Mass (\Msun)       & 1.68$\pm$0.25      & 1                    & 1.66$\pm$0.25  & 1                      & 1.07$\pm$0.25   & 1                             \\
Distance (pc)      & 45.1$\pm$0.5      & 4                     & 169$\pm$22   & 4                        & 83.5$\pm$4.3    & 4                             \\
$[Fe/H]$           & 0.07$\pm$0.10     & 1                     & 0.04$\pm$0.10  & 1                      & 0.05$\pm$0.10   & 1                             \\
                   & 0.11$\pm$0.03     & 3                     & 0.08$\pm$0.09  & 7                      & 0.16            & 6                             \\
T$_{\rm{eff}}$ (K) & 4848$\pm$100      & 1                     & 4805$\pm$100   & 1                      & 4806$\pm$100    & 1                             \\
                   & 4828$\pm$53       & 3                                                               & 4900            & 6                             \\
                   & 5170              & 5                     & 4861         & 5                        & 4950            & 5                             \\
log $g$            & 3.33$\pm$0.15     & 1                     & 2.66$\pm$0.15  & 1                      & 3.31$\pm$0.15   & 1                             \\
                   & 3.27$\pm$0.16     & 3                                                               & 3.18            & 6                             \\
Luminosity (\Lsun) & 10.0              & 1                     & 36.3          & 1                       & 8.5             & 1                             \\
                   & 9.56              & 5                     & 33.38         & 5                       & 7.84            & 5                             \\
Radius (\Rsun)     & 4.0$\pm$0.6      & 1                     & 8.8$\pm$0.6    & 1                      & 4.1$\pm$0.6     & 1                             \\
\enddata
\tablecomments{References: 1 - \citet{ppps5}, 2 - \citet{houk75}, 3 - 
\citet{alves15}, 4 - \citet{vl07}, 5 - \citet{Macdonald2012}, 6 - 
\citet{jones11}, 7 - \citet{houk88}, 8 - \citet{landolt83}, 9 - 
\citet{corben72}, 10 - \citet{hog00} }
\end{deluxetable}


\begin{deluxetable}{lrr}
\tabletypesize{\scriptsize}
\tablecolumns{3}
\tablewidth{0pt}
\tablecaption{Keplerian orbital solutions}
\tablehead{
\colhead{Parameter} & \colhead{HD 86950b} & \colhead{HD 222076b}}
\startdata
\label{planetparams}
Period (days) & 1270$\pm$57 & 871$\pm$19 \\
$T_0$ (BJD-2400000) & 54245$\pm$161 & 54264$\pm$189 \\ 
Eccentricity &  0.17$\pm$0.16 & 0.08$\pm$0.05 \\
$\omega$ (degrees) &  243$\pm$70 & 241$\pm$60 \\
$K$ (\ms) &  49$\pm$12 & 31.9$\pm$2.3 \\ 
m sin $i$ (\Mjup) &  3.6$\pm$0.7 & 1.56$\pm$0.11 \\
$a$ (AU) &  2.72$\pm$0.08 & 1.83$\pm$0.03 \\
RMS about fit (\ms) &  6.1 & 5.9 \\
\hline
Zero point -- AAT \ms\ & 13.9$\pm$4.7 & -1.6$\pm$2.1 \\
Zero point -- FEROS \ms\ & \nodata & -10.3$\pm$1.8 \\
Zero point -- CHIRON \ms\ & -17.8$\pm$20.4 & -10.3$\pm$5.3 \\
\enddata
\end{deluxetable}


\begin{deluxetable}{llll}
\tabletypesize{\scriptsize}
\tablecolumns{4}
\tablewidth{0pt}
\tablecaption{FEROS activity indicators for HD\,222076}
\label{ferosact}
\tablehead{
\colhead{BJD-2400000} & \colhead{BVS (\ms)} & \colhead{FWHM (\kms)} & 
\colhead{$S_{MW}$} }
\startdata
55317.8897   &   0.6$\pm$10.9  & 12.749$\pm$0.065  & 0.138$\pm$0.004 \\
55379.9038   &  -6.1$\pm$7.3  & 12.767$\pm$0.076  & 0.131$\pm$0.002 \\
55428.7555   & -20.4$\pm$12.4  & 12.753$\pm$0.064  & 0.125$\pm$0.002 \\
55457.7279   & -24.1$\pm$7.9  & 12.855$\pm$0.077  & 0.127$\pm$0.002 \\
55729.9102   &  -5.4$\pm$8.4  & 12.854$\pm$0.072  & 0.125$\pm$0.003 \\
55786.8991   & -12.0$\pm$9.8  & 12.841$\pm$0.076  & 0.134$\pm$0.002 \\
55793.8343   & -29.2$\pm$8.3  & 12.848$\pm$0.075  & 0.133$\pm$0.003 \\
56047.9343   &  17.2$\pm$6.6  & 12.856$\pm$0.077  & 0.128$\pm$0.002 \\
56160.7983   & -39.0$\pm$9.4  & 12.835$\pm$0.072  & 0.126$\pm$0.002 \\
56241.6291   & -22.8$\pm$13.6  & 12.767$\pm$0.063  & 0.131$\pm$0.002 \\
56251.6645   &   1.5$\pm$12.7  & 12.832$\pm$0.072  & 0.117$\pm$0.002 \\
56412.8628   &  -1.0$\pm$10.5  & 12.716$\pm$0.067  & 0.093$\pm$0.004 \\
56431.8403   &  -7.7$\pm$7.5  & 12.793$\pm$0.066  & 0.137$\pm$0.004 \\
56565.6838   & -10.2$\pm$12.4  & 12.811$\pm$0.066  & 0.122$\pm$0.002 \\
57174.8723   &  -2.7$\pm$7.0  & 12.780$\pm$0.068  & 0.126$\pm$0.002 \\
57388.5551   & -32.4$\pm$11.2  & 12.825$\pm$0.072  & 0.132$\pm$0.002 \\
57389.5509   & -20.7$\pm$13.6  & 12.738$\pm$0.065  & 0.125$\pm$0.002 \\
\enddata
\tablecomments{BVS -- bisector velocity span; FWHM -- full-width half max of tghe 
cross-correlation function; $S_{MW}$ -- Mount Wilson S-index.}
\end{deluxetable}

\clearpage

\clearpage

\clearpage

\clearpage

\clearpage

\clearpage

\begin{deluxetable}{lllll}
\tabletypesize{\scriptsize}
\tablecolumns{5}
\tablewidth{0pt}
\tablecaption{Detection fraction in host star mass/metallicity bins. Bin 
sizes are the same as in \citet{reffert15}, their Table 1. }
\tablehead{
\colhead{[Fe/H]} & \colhead{$M_*$ (\Msun)} & \colhead{$N_{stars}$} & 
\colhead{$N_{hosts}$} & \colhead{$f$ (\%)} }
\startdata
\label{metal}
-0.36 &  1.4  &   10   &   0   &    $0.0^{+9.1}_{-0.0}$  \\
-0.36 &  2.2  &   1    &   0   &    $0.0^{+50.0}_{-0.0}$  \\   
-0.36 &  3.0  &   0    &   0   &    \nodata \\
-0.20 &  1.4  &   27   &   1   &    $3.7^{+5.6}_{-2.3}$  \\ 
-0.20 &  2.2  &   2    &   0   &    $0.0^{+33.4}_{-0.0}$  \\ 
-0.20 &  3.0  &   0    &   0   &    \nodata \\
-0.04 &  1.4  &   60   &   4   &    $6.7^{+4.0}_{-2.6}$  \\ 
-0.04 &  2.2  &   4    &   1   &    $25.0^{+25.0}_{-15.0}$  \\
-0.04 &  3.0  &   0    &   0   &    \nodata \\
+0.12 &  1.4  &   49   &   5   &    $10.2^{+5.2}_{-3.6}$  \\ 
+0.12 &  2.2  &   0    &   0   &    \nodata \\
+0.12 &  3.0  &   0    &   0   &    \nodata \\
+0.28 &  1.4  &   10   &   0   &    $0.0^{+9.1}_{-0.0}$  \\ 
+0.28 &  2.2  &   0    &   0   &    \nodata \\
+0.28 &  3.0  &   0    &   0   &    \nodata \\
\enddata
\end{deluxetable}

\clearpage

\clearpage

\appendix

\section{Continuum normalization}

In our UCLES data, the H$\alpha$ line is recorded in two echelle 
orders (orders 19 and 20).  Our analysis makes use of only the data in 
order 19.  Order 20 (which is closer to the edge of the echellogram 
format) was found to be more subject to scattered light contamination 
than order 19, and so delivered systematically poorer normalization. The 
order 19 spectra were analysed in a $10\mbox{\AA}$ window centred on the 
H$\alpha$ line.

Our continuum fitting algorithm iteratively masks out pixels in 
the chosen wavelength range based on (1) their first-derivative (i.e. 
rejecting pixels until no neighbouring pixel has a gradient slopes 
larger than 10\% of the maximum gradient of the original spectrum 
section), after which it (2) removes the pixels ``left behind'' at the 
bottoms of absorption features by keeping the local maxima and removing 
the local minima.  The remaining unmasked continuum points are then (3) 
fit with linear regression, which is divided out of the original 
spectrum $S_0$ to create a first-pass normalized spectrum $s_0$.  Cosmic 
rays are removed from this first-pass continuum-normalised spectrum by 
rejecting pixels more than 10\% above the normalised continuum (i.e. 
with values $s_0>1.1$).  Steps (1)-(3) are then repeated using the 
original spectrum $S_0$ but with the cosmic ray pixels (i.e. the 
outliers above the continuum) removed to produce a final 
continuum-normalized spectrum $s_1$. 

We then construct a high $S/N$ H$\alpha$ template and with it we 
remove further cosmic rays from the H$\alpha$ window (labelled within 
dashed lines in the stacked spectra) by identifying groups of pixels 
displaying anomalous variability between exposures.  We select the 
normalized spectrum with the highest $S/N$ as a first template $T_1$, 
then shift all other spectra to match the velocity of $T_1$, using the 
velocities of \S\ref{obs}, to create a high $S/N$ template $T_2$ as the 
weighted average of the individual spectra.  We reject pixels in the 
H$\alpha$ window of individual spectrum that combine to form $T_2$ which 
deviate from the normalized mean by more than empirically 21 times the 
expected noise to signal ratio. 

The measurement of H$\alpha$ activity indices can be impacted by 
the presence of telluric absorption lines.  Unfortunately, B star 
observations of each night are not available for telluric correction.  
We therefore generate a list of telluric line locations using the HITRAN 
(HIgh-resolution TRANsmission molecular absorption) database 
\citep{Rothman2013}, and use this to mask out the regions potentially 
contaminated by $\mathrm{H_2O}$ absorption.  In the specific case of our 
data the prominent line is $6564.258\mbox{\AA}$ at rest frame.  The 
telluric contamination is labelled in the shaded region of the stacked 
profile. 

To avoid the telluric contamination, we define our H$\alpha$ 
equivalent width as the sum of the normalised flux obtained from above 
at a $0.5\mbox{\AA}$ window centred at H$\alpha$ line 
$\sum_{|\lambda-\lambda_{\mathrm{H}\alpha}|<0.5\AA} \big(1 - 
F_\lambda\big)$.


\end{document}